\documentclass[a4paper,fleqn,usenatbib]{mnras}

\usepackage{newtxtext,newtxmath}

\usepackage[T1]{fontenc}
\usepackage{ae,aecompl}

\usepackage{changes}
\colorlet{Changes@Color}{red}

\usepackage{graphicx}	
\usepackage{amsmath}	
\usepackage{amssymb}	
 \usepackage{adjustbox}
 \usepackage{ulem}
 
\usepackage{xcolor}
 
\usepackage{cancel}
 
 \usepackage{subfig}
 
\title[Identifying \HII\ regions from IRAS PSC]{Improved selection criteria for \HII\ regions, based on IRAS sources}

\author[Q. Z. Yan et al.]{Qing-Zeng Yan,$^{1,2,3,4}$\thanks{E-mail: qzyan@shao.ac.cn}
Ye Xu,$^{2}$\thanks{xuye@pmo.ac.cn}
A. J. Walsh,$^{3}$
J. P. Macquart,$^{3}$   
G. C. MacLeod,$^{5}$ \and
Bo Zhang,$^{1}$ 
P. J. Hancock,$^{3}$  
Xi Chen,$^{6,1}$\thanks{chenxi@shao.ac.cn}
Zheng-Hong Tang,$^{1,7}$\\  
\\
$^{1}$ Shanghai Astronomical Observatory, Chinese Academy of Sciences, Shanghai 200030, China\\
$^{2}$ Purple Mountain  Observatory, Chinese Academy of Sciences, Nanjing 210008, China \\
$^{3}$ International Centre for Radio Astronomy Research, Curtin University, GPO Box U1987, Perth WA 6845, Australia\\
$^{4}$ University of Chinese Academy of Sciences, 19A Yuquanlu, Beijing 100049, China\\
$^{5}$ Hartebeesthoek Radio Astronomy Observatory, P.O. Box 443, Krugersdorp 1740, South Africa\\
$^{6}$ Center for Astrophysics, GuangZhou University, Guangzhou 510006, China;\\
$^{7}$ School of Astronomy and Space Science, University of Chinese Academy of Sciences, 19A Yuquanlu, Beijing 100049, China\\
}
\date{Accepted 2018 Feb 20}

\pubyear{2018}

\def\deg{$\degr$}

\newcommand{\HII}{\mbox{H\,\textsc{ii}}}%
\newcommand{\um}{$\mathrm{\mu}$m}

\newcommand{\wise}{\emph{WISE}}

\newcommand{\Fone}{$F_{12}$}
\newcommand{\Ftwo}{$F_{25}$}
\newcommand{\Fthree}{$F_{60}$}
\newcommand{\Ffour}{$F_{100}$}

\RequirePackage[normalem]{ulem} 
\RequirePackage{color}\definecolor{RED}{rgb}{1,0,0}\definecolor{BLUE}{rgb}{0,0,1} 

\definecolor{cadmiumgreen}{rgb}{0.0, 0.42, 0.24}
\definecolor{mygreen}{rgb}{0.0, 0.5, 0.0}
\begin{document}
\label{firstpage}
\pagerange{\pageref{firstpage}--\pageref{lastpage}}
\maketitle

\begin{abstract}
We present new criteria for selecting \HII\ regions from the Infrared Astronomical Satellite (IRAS) Point Source catalogue (PSC), based on an \HII\ region catalogue derived manually from the all-sky \emph{Wide-field Infrared Survey Explorer} (\emph{WISE}). The criteria are used to augment the number of \HII\   region candidates in the Milky Way. The criteria are defined by the linear decision boundary of two samples: IRAS point sources associated with known  \HII\ regions, which serve as the \HII\ region sample, and IRAS point sources at high Galactic latitudes, which serve as the non-\HII\ region sample. A machine learning classifier, specifically a support vector machine (SVM), is used to determine the decision boundary. We investigate all combinations of four IRAS bands and suggest that the optimal criterion is $\mathrm{log}\left(F_{60}/F_{12}\right)\gid \left(  -0.19 \times \mathrm{log}\left(F_{100}/F_{25}\right)+ 1.52\right)$, with detections at 60 and 100 \um. This  selects 3041 \HII\ region candidates from the IRAS PSC. We find that IRAS \HII\ region candidates show evidence of evolution on the two-colour diagram. Merging the \emph{WISE} \HII\ catalogue with IRAS \HII\ region candidates, we estimate  a lower limit of approximately 10200 for the number of \HII\ regions in the Milky Way. 


\end{abstract}




\section{Introduction}
\label{sec:intro}

High-mass stars, whose masses exceed 8 $M_\odot$~\citep{2007ARA&A..45..481Z}, are OB stars which emit strong UV radiation, thereby ionizing surrounding atomic and molecular gases. Composed mainly of ionized hydrogen, these ionized gases are usually called \HII\ regions. Most  \HII\ regions essentially trace high-mass stars, hence the evolution and distribution of \HII\ regions is useful for investigating high-mass stars in the Milky Way. 

 
Although the amount of \HII\ regions  is an important evolutionary indicator of the Milky Way, their total number is still unclear due to the difficulty in detecting and identifying them due to their large distances and rapid evolution~\citep{2007ARA&A..45..481Z}. In their early stages, when \HII\ regions are compact, high-mass stars are deeply embedded in cold thick molecular clouds, whose typical temperature is about 30 K~\citep{1994ApJ...427..889W,1999PASP..111.1049G}. In this phase, high-mass stars are invisible at optical wavelengths due to high extinction caused by dust grains, but they are observable at longer wavelengths, notably in the infrared or radio bands. The infrared output of \HII\ regions is generally due to the thermal emission of their internal or surrounding dust grains~\citep{1999PASP..111.1049G,2002ARA&A..40...27C}, whereas the radio output is generated by their internal free-free emission~\citep{1994ApJS...91..659K,1998MNRAS.301..640W}. However, the properties of infrared emission vary with temperature, while the optical depth of free-free emission hinges on the frequency~\citep{1994ApJS...91..659K}.





\citet[here after WC89]{1989ApJ...340..265W} investigated the population and distribution of embedded high-mass stars in the Milky Way using the all-sky Infrared Astronomical Satellite (IRAS) Point Source catalogue (PSC)~\citep{1984ApJ...278L...1N, 1988iras....7.....H} over four infrared bands at 12 \um, 25 \um, 60 \um\ and 100 \um.  If there is no detection in one band, the flux density quality of this band is marked with an upper limit. With the help of some previously identified Ultra-Compact (UC)  \HII\ regions~\citep{1989ApJS...69..831W}, they derived a criterion for embedded high-mass stars: $\mathrm{log}(F_{60}/F_{12})\gid  1.30$ and $\mathrm{log}(F_{25}/F_{12})\gid 0.57$, where \Fone, \Ftwo, and \Fthree\ represent fluxes at  12 \um, 25 \um, 60 \um, respectively. They further rejected sources whose flux density quality at either 25 or 60 \um\ is marked by an upper limit. They identified 1717 UC \HII\ region candidates and  potentially missed many evolved \HII\ regions.


 
Using known \HII\ regions, including UC \HII\ regions,  \citet[here after HM89]{1989AJ.....97..786H}  investigated the \HII\ regions based on a two-colour diagram of IRAS sources.  They provided a decision boundary of $\mathrm{log}(F_{25}/F_{12})\gid 0$ and  $\mathrm{log}(F_{60}/F_{25})\gid0$. They also imposed extra constraints on $F_{100}$ and Galactic latitudes, and the total number of \HII\   region candidates identified was 2298. However,  the sample of known \HII\ regions they used is far from complete, and therefore the criterion were not well constrained.

Recently, \citet{2014ApJS..212....1A} created a catalogue of \HII\  region candidates, providing an opportunity to improve the selection criteria for  \HII\  region candidates. This \HII\ region  candidate catalogue is based on the all-sky \emph{Wide-field Infrared Survey Explorer} (\emph{WISE})~\citep{2010AJ....140.1868W}. They created this catalogue by identifying  infrared bubbles~\citep{2006ApJ...649..759C} manually.  Infrared bubbles essentially are \HII\ regions produced by high-mass stars. The 12 \um\ band emission traces polycyclic aromatic hydrocarbon (PAH) molecules, delineating the edge of \HII\ regions, while the 24  \um\ band emission mostly traces internal thermal emission from dust grains heated by ionized gases. This is the most complete catalogue of \HII\ regions in the Milky Way, because the covering area is much larger than that of the \emph{Spitzer}/GLIMPSE survey~\citep{2003PASP..115..953B}, and the number (8399) of identified \HII\ regions  exceeds that provided by another similar undertaking: the Milky Way Project (MWP)~\citep{2012MNRAS.424.2442S}.

 However, the catalogue of \citet{2014ApJS..212....1A} can potentially miss those \HII\ regions that have small angular  sizes or that are not easily identified visually. In this paper, we investigate possible criteria for selecting \HII\ regions from IRAS sources to obtain a more complete census of \HII\ regions  in the Milky Way, using publicly available radio and infrared surveys and sophisticated algorithms. Essentially, the criteria are defined by the decision boundary of two samples: IRAS sources which are associated  known  \HII\ regions and those sources which are  not \HII\ regions.  The \emph{WISE} \HII\ region candidates~\citep{2014ApJS..212....1A} are used as a basis for the \HII\ region sample (subject to further selection criteria based on radio detection), while the IRAS point sources at high Galactic latitudes serve as the non-\HII\ regions. Support vector machines (SVMs)~\citep{vapnik1995nature}, which are machine learning algorithms used to do supervised classification, are applied to derive the decision boundary of the two samples.  A 3D simulation of the expansion of \HII\ regions~\citep{2014A&A...568A...4T} enables us to estimate the age of \HII\ regions and to investigate the evolution of \HII\ regions on two-colour diagrams of IRAS sources.

\section{ANALYSIS} \label{sec:analysis}


In this section, we present our method of producing selection criteria for \HII\ regions, based on the IRAS PSC and the \wise\ \HII\ region catalogue. Essentially, the criteria are determined by two types of IRAS point sources: \HII\ regions, and non-\HII\ regions. We match the IRAS PSC to the positions of the \wise\ \HII\ region  candidates that have radio counterparts, yielding the \HII\ region sample, whereas the non-\HII\ region sample is built from high Galactic latitude sources ($|b|>8\degr$).  The decision boundary of the two samples is produced by the SVM algorithm and is subsequently applied to the IRAS PSC to identify \HII\ region candidates.

\subsection{catalogues}

Our analysis is based on the IRAS PSC~\citep{1988iras....7.....H}, the \wise\ \HII\ region catalogue~\citep{2014ApJS..212....1A}, and three radio continuum source catalogues~\citep{1998AJ....115.1693C,2003MNRAS.342.1117M,2007MNRAS.382..382M}.

The IRAS PSC (version 2.1) includes four infrared fluxes at 12, 25, 60, and 100 \um, with the resolution ranging from $45\arcsec$ to $3\arcmin$. We use \Fone, \Ftwo, \Fthree\ and \Ffour\ to denote fluxes at  these four bands, respectively, and use $Q_{12}$, $Q_{25}$, $Q_{60}$ and $Q_{100}$ to denote their corresponding qualities. The flux quality values of 1, 2, and 3 represent an upper limit (i.e. non-detection), moderate quality, and high quality, respectively. After eliminating galaxies and quasars identified by~\citet{1989IRASG....c...0F}, we further rejected those IRAS sources which are matched with extragalactic objects within $1\arcmin$, including nearby galaxies~\citep{1986A&AS...66..255K, 2013AJ....145..101K, 2015ApJS..220....6B} and unresolved very long baseline interferometry (VLBI) calibrators identified by \citet{2006ApJS..166..526X} and \citet{2011ApJS..194...25I}. Our study is based on the remaining 234261 IRAS point sources.

The \HII\ region catalogue that we use to extract \HII\ regions from the IRAS PSC is created by~\citet{2014ApJS..212....1A} from \wise\ data. This catalogue contains 8399 \HII\  region candidates, covering $|b|\lid8\degr$ and five high-mass star-forming regions at high Galactic latitudes.  In this catalogue, 1413 \HII\ regions have their distances determined,  and we estimate their ages based on a numerical simulation of the 3D expansion of \HII\  region \citep{2014A&A...568A...4T}.

  
In their catalogue, \citet{2011ApJS..194...32A} found that the coexistence of radio continuum and mid-infrared emission can identify \HII\ regions at a 95\% confidence level. Therefore, in order to examine the quality of \HII\ region candidates, we use three radio continuum surveys: the NRAO VLA Sky Survey (NVSS)~\citep{1998AJ....115.1693C}, the Sydney University Molonglo Sky Survey (SUMSS)~\citep{1999AJ....117.1578B, 2003MNRAS.342.1117M}, and the second epoch Molonglo Galactic Plane Survey (MGPS-2)~\citep{2007MNRAS.382..382M}.  NVSS, SUMSS, and MGPS-2 have similar sensitivities and spatial resolutions and collectively cover the whole sky. At frequencies $\nu\lid$ 8-15 GHz, radio free-free emission is optically thick~\citep{1994ApJS...91..659K,2008MNRAS.390..683P} for UC \HII\ regions, rendering UC \HII\ regions undetectable at these frequencies. Nonetheless, the proportion of sources associated with radio continuum emission is still an excellent indicator of the quality of the selection criteria, because only those \HII\ regions at very early stages are missed.



We summarize the five catalogues in Table~\ref{tab:surveys}, where from left to right,  we list the name, the telescope, the observed band, the spatial resolution,  the covering area, and the reference for each catalogue.

 \begin{table*}
	\centering
\caption{catalogues of IRAS point sources, \HII\ regions, and three radio continuum surveys. }
\label{tab:surveys}
 	\begin{tabular}{cccccc}
 	\hline
 catalogue &  Telescope  &  Band & Resolution  & Coverage   & Reference \\
\hline 
IRAS v2.1 &IRAS &  12, 25, 60, and 100 \um  & $45''-3'$ &  all-sky  &   \citet{1984ApJ...278L...1N,1988iras....7.....H} \\
 
\HII\ regions &\wise &  12 and 22 \um  & $ 6\farcs5$ and $12\arcsec$ &  $|b|\lid8\degr$  &   \citet{2014ApJS..212....1A} \\
 
NVSS &VLA& 1.4 GHz & 45$''$ & $\delta >-40^\circ$ &   \citet{1998AJ....115.1693C} \\
 
SUMSS &Molonglo& 843 MHz & $\sim$45$''$ &  $\delta<-30^\circ(|b|>10^\circ) $&   \citet{1999AJ....117.1578B, 2003MNRAS.342.1117M} \\
 
MGPS-2 &Molonglo& 843 MHz & $\sim$45$''$ &  $245^\circ<l< 365^\circ$, $|b|<10^\circ$ &   \citet{2007MNRAS.382..382M} \\
\hline
	\end{tabular}
\end{table*}


 \begin{table*} 
	\centering

\caption{All possible two-colour criteria for selecting \HII\ regions from the IRAS PSC.  \label{tab:criteria}}
        \begin{adjustbox}{max width=0.99\textwidth,center}
          
 	\begin{tabular}{clccccccc}
 	\hline
 Identity  & Criterion $\rm ^c$ &   Candidates &  Association$\rm^d$  &  Sensitivity$\rm^e$  &  Precision$\rm ^f$ &  Tested$\rm^g$&  Score$\rm^h$\\
 &  &    & (Radio)  &    &     &  \\
  	\hline  
1$\rm^a$ &$\mathrm{log}\left(\frac{F_{60}}{F_{12}}\right)>\left( -0.19 \times \mathrm{log}\left(\frac{F_{100}}{F_{25}}\right)+1.52\right)$, $Q_{60} \ \& \ Q_{100} >1$     & 3041 & 32.46\% & 90.58\% &93.32\% & 24441 &0.5707  \\  
 2 &$\mathrm{log}\left(\frac{F_{100}}{F_{60}}\right)>\left( -3.43 \times \mathrm{log}\left(\frac{F_{60}}{F_{12}}\right)+4.73\right)$, $Q_{60} \ \& \ Q_{100} >1$     & 3077 & 32.11\% & 90.48\% &93.70\% & 24441 &0.5674  \\  
 3 &$\mathrm{log}\left(\frac{F_{60}}{F_{12}}\right)>\left( -0.45 \times \mathrm{log}\left(\frac{F_{100}}{F_{12}}\right)+2.00\right)$, $Q_{60} \ \& \ Q_{100} >1$     & 3126 & 31.70\% & 90.48\% &93.41\% & 24441 &0.5628  \\  
 4$\rm^b$ &$\mathrm{log}\left(\frac{F_{60}}{F_{100}}\right)>\left( -1.47 \times \mathrm{log}\left(\frac{F_{100}}{F_{12}}\right)+2.03\right)$, $Q_{60} \ \& \ Q_{100} >1$     & 3128 & 31.55\% & 90.48\% &93.31\% & 24441 &0.5612  \\  
 5 &$\mathrm{log}\left(\frac{F_{60}}{F_{12}}\right)>\left( 0.19 \times \mathrm{log}\left(\frac{F_{25}}{F_{12}}\right)+1.18\right)$, $Q_{25} \ \& \ Q_{60} >1$     & 4502 & 30.16\% & 89.01\% &92.77\% & 24941 &0.5438  \\  
 6 &$\mathrm{log}\left(\frac{F_{25}}{F_{12}}\right)>\left( -1.25 \times \mathrm{log}\left(\frac{F_{60}}{F_{25}}\right)+1.47\right)$, $Q_{25} \ \& \ Q_{60} >1$     & 4517 & 30.04\% & 89.17\% &92.71\% & 24941 &0.5426  \\  
 7 &$\mathrm{log}\left(\frac{F_{60}}{F_{25}}\right)>\left( -5.53 \times \mathrm{log}\left(\frac{F_{60}}{F_{12}}\right)+7.90\right)$, $Q_{60}>1$& 4859 & 29.29\% & 86.11\% &92.57\% & 60315 &0.5304  \\  
 8 &$\mathrm{log}\left(\frac{F_{100}}{F_{12}}\right)>\left( 0.41 \times \mathrm{log}\left(\frac{F_{100}}{F_{25}}\right)+1.18\right)$, $Q_{100}>1$& 3742 & 27.15\% & 80.20\% &91.26\% & 60349 &0.4979  \\  
 9 &$\mathrm{log}\left(\frac{F_{100}}{F_{60}}\right)>\left( -1.47 \times \mathrm{log}\left(\frac{F_{25}}{F_{12}}\right)+0.86\right)$, $Q_{25} \ \& \ Q_{100} >1$     & 3599 & 28.59\% & 77.62\% &78.88\% & 10714 &0.4956  \\  
 10 &$\mathrm{log}\left(\frac{F_{100}}{F_{12}}\right)>\left( -0.31 \times \mathrm{log}\left(\frac{F_{25}}{F_{12}}\right)+1.66\right)$, $Q_{25} \ \& \ Q_{100} >1$     & 4439 & 25.21\% & 92.41\% &91.77\% & 10714 &0.4887  \\  
 11 &$\mathrm{log}\left(\frac{F_{100}}{F_{25}}\right)>\left( -1.29 \times \mathrm{log}\left(\frac{F_{25}}{F_{12}}\right)+1.65\right)$, $Q_{25} \ \& \ Q_{100} >1$     & 4491 & 25.01\% & 93.01\% &91.54\% & 10714 &0.4865  \\  
 12 &$\mathrm{log}\left(\frac{F_{60}}{F_{25}}\right)>\left( -1.83 \times \mathrm{log}\left(\frac{F_{100}}{F_{12}}\right)+4.01\right)$, $Q_{60} \ \& \ Q_{100} >1$     & 4111 & 22.70\% & 85.41\% &88.36\% & 24441 &0.4472  \\  
 13 &$\mathrm{log}\left(\frac{F_{60}}{F_{25}}\right)>\left( -0.02 \times \mathrm{log}\left(\frac{F_{100}}{F_{60}}\right)+1.02\right)$, $Q_{60} \ \& \ Q_{100} >1$     & 3083 & 25.43\% & 60.08\% &79.07\% & 24441 &0.4372  \\  
 14 &$\mathrm{log}\left(\frac{F_{60}}{F_{25}}\right)>\left( -0.03 \times \mathrm{log}\left(\frac{F_{100}}{F_{25}}\right)+1.06\right)$, $Q_{60} \ \& \ Q_{100} >1$     & 3099 & 25.20\% & 59.78\% &79.19\% & 24441 &0.4346  \\  
 15 &$\mathrm{log}\left(\frac{F_{100}}{F_{25}}\right)>\left( 0.96 \times \mathrm{log}\left(\frac{F_{100}}{F_{60}}\right)+1.05\right)$, $Q_{100}>1$& 4600 & 19.65\% & 58.09\% &78.30\% & 60349 &0.3710  \\  
  	\hline
 
 \multicolumn{8}{l}{$\rm^a$ Criterion 1 is the optimal criterion.}\\

 \multicolumn{8}{l}{$\rm^b$ In order to make sure \HII\ regions are above the decision boundary, we use $F_{60}/F_{100}$ instead of $F_{100}/F_{60}$ for criterion 4.}\\
 \multicolumn{8}{l}{$\rm^c$ The slopes and intercepts have been rounded up, the error caused by which is not significant.}\\
 \multicolumn{8}{l}{$\rm^d$ The proportion of candidates possessing radio counterparts (radio association).}\\
 \multicolumn{8}{l}{$\rm^e$ The sensitivity is defined in Equation~\ref{eq:measures}.}\\
 \multicolumn{8}{l}{$\rm^f$ The precision is defined in Equation~\ref{eq:measures}.}\\
 \multicolumn{8}{l}{$\rm^g$ The number of IRAS sources tested by the inequality of criteria.}\\
 \multicolumn{8}{l}{$\rm^h$ The score is defined by equation \ref{eq:score}. }\\

	\end{tabular}
\end{adjustbox}
\end{table*}

 \subsection{Support vector machines}

 The SVM algorithm is used to decide the decision boundary of \HII\ regions. SVMs, developed by Vladimir Vapnik~\citep{vapnik1995nature, cortes1995support}, are algorithms used to do classification and regression analysis in supervised machine learning. For two groups of points, which are linearly separable, SVMs determine their decision boundary by maximizing the gap between them. However, if they are not linearly separable, SVMs can still perform classification by mapping them into a higher dimensional space using specific kernels.

We adopted linear SVM classifiers, because the overlapping area of \HII\ regions and non-\HII\ regions is not large, meaning they are well separated.  We use the \emph{Python} package \emph{sklearn} to perform linear SVMs. An important parameter of this algorithm is the penalty parameter for misclassification, denoted by \emph{C}. Larger values of \emph{C} impose higher penalties for misclassification, while smaller values of \emph{C} permit more misclassification.

Due to the presence of overlapping areas, we allow a small fraction of misclassification, but the misclassification needs to be firmly constrained to avoid involving a larger number of high Galactic latitude sources. Consequently, we adopt an intermediate value $\emph{C}=1$. In \S\ref{sec:dosvm}, we find the effect caused by a small shift of \emph{C} is not significant.



 \subsection{Producing the criteria}
 \label{sec:dosvm}
   
HM89 investigated the selection criteria for \HII\ regions based on an incomplete sample of \HII\ regions, while we use a more complete \HII\ region catalogue and more sophisticated algorithms to improve their result in this subsection. The entire process of creating selection criteria is divided into five main steps, details of which are described in the rest of this subsection. We choose the optimal criterion  according to a metric, which is defined below (in Equation~\ref{eq:score}). 
  
 The process of producing criteria includes five main steps: 


\begin{enumerate}
\item Identify IRAS sources associated with  \HII\ regions, which serve as the sample of known \HII\ regions.
 \item Select  IRAS sources at high Galactic latitudes, which serve as the sample of non-\HII\ regions.
 \item Use SVMs to determine the criteria based on the samples of \HII\ regions and non-\HII\ regions for all possible colour combinations.

 \item  Apply these criteria to the IRAS PSC.
  \item  Determine the optimal criterion according to their scores.
\end{enumerate}



In the first step, we matched the IRAS PSC to the \wise\ \HII\ region catalogue.  Following~\citet{2014ApJS..212....1A}, we only adopted \wise\ sources that have small angular sizes (radii $<4\arcmin$) and ignore those lack detected radio continuum emission (the classification ``Q''). In total, 1773 IRAS sources are matched with at least one \HII\ region, and these IRAS sources serve as the sample of \HII\ regions. For a particular colour combination, these sources are  further filtered with bands required to have good qualities (better than an upper limit).  



The second step is to build the sample of non-\HII\ regions, by selecting IRAS sources at high Galactic latitudes. This is because high-mass stars are generally far away from the Sun and are tightly constrained to the Galactic plane~\citep{2007ARA&A..45..481Z}. Furthermore, they are not overlapping with any of the giant molecular clouds (GMCs) in the Milky Way identified by~\citet{2016ApJ...822...52R}. At high Galactic latitudes (except the Orion nebula, which is a high-mass star forming region at a distance of about 400 pc from the Sun), most IRAS point sources are low-mass stars or extragalactic objects. Although some  extragalactic objects possess similar colours to \HII\ regions, there are relatively few, on account of  the reddening caused by the intergalactic medium (IGM)~\citep{1981ApJ...250....1W, 2013ApJ...772...26A}. 

Because the \wise\ \HII\ region catalogue extends up to a Galactic latitude of 8\degr, we selected those IRAS sources whose absolute values of Galactic latitude are greater than 8$^\circ$ ($|b|>8\degr$), serving as the sample of non-\HII\ regions. We further rejected those sources in three prominent regions: Orion, the Large Magellanic Cloud (LMC), and the Small Magellanic Cloud (SMC). According to a CO survey performed by~\citet{2005A&A...430..523W}, the range of Galactic longitude and latitude for Orion are [200\deg, 220\deg] and [-22\deg, -8\deg], respectively. Based on the position and size of galaxies provided by~\citet{2014MNRAS.445..881C}, the Galactic longitude ranges of the LMC and the SMC are  [275\deg, 286\deg] and  [  299\deg, 305.5\deg] and their Galactic latitude ranges are [-38.5\deg, -27.5\deg] and  [-47\deg, -41.5\deg], respectively. 

We used the criteria of WC89 and HM89 (no constraints on Galactic latitudes) to examine the quality of those sources at high Galactic latitudes. In total, 6200 sources are tested by the inequalities of WC89 or HM89, and we find 89 (1.4\%) sources at high Galactic latitude ($|b|>8\degr$) agree with at least one of these two criteria. After eliminating these 89 sources, we have remaining  100634  sources at high Galactic latitudes, which is used to build the sample of non-\HII\ regions. Despite the incompleteness of WC89 and HM89,  we estimate that less than 1.4\% of those sources are possibly \HII\ regions.



In the third step, we checked all the possible two-colour combinations of IRAS bands,  each of which involves at least three bands. For each colour, we require that the shorter wavelength is the denominator so that \HII\ regions will be above the decision boundary. For an IRAS point source, if the shorter wavelength band of its colour is marked with an upper limit, the true value will only move this source to the upper right direction on the two-colour diagram.   Therefore, we only require that the flux quality at longer wavelengths is better than an upper limit. In total, there are 15 possible two-colour combinations. With the help of SVMs, we determined the decision boundary for each colour combination, and as mentioned above, the penalty parameter (\emph{C}) is assigned a value of 1, and the effect caused by a small change of \emph{C} is negligible.  Generally, within the range $0.5 <$\emph{C}$<1.5$, the shifts of slopes and y-intercepts of the  criteria are less than 0.01. 
 
Before we performed the fourth step, we calculated some statistical measures for the criteria, including sensitivity, specificity, and the $F_1$ score. The true positives ($TP$) are the \HII\ region samples agreeing with the criteria, while the false negatives ($FN$) are the \HII\ region samples rejected by the criteria. The true negatives ($TN$) are the non-\HII\ region samples rejected by the criteria, while the false positives ($FP$) are the non-\HII\ region samples agreeing with the criteria. The definition of sensitivity, precision, and the $F_1$ score are

 \begin{equation}
 \label{eq:measures}
   \begin{array}{ccc}
      \mathrm{sensitivity}& =&\frac{TP}{\left(TP+FN\right)}, \\ 
      \mathrm{precision}& =&\frac{TP}{\left(TP+FP\right)},\\                 
     F_1\,\mathrm{score}& =&\frac{2}{\left(1/\mathrm{sensitivity}+1/\mathrm{precision}\right)},\\ 
   \end{array}
\end{equation}
where sensitivity is also called the recall or the true positive rate (TPR), precision is also called the positive predictive value (PPV), and the $F_1 \, \mathrm{score}$ is the harmonic mean of precision and sensitivity. In the fourth step, we modify the $F_1 \, \mathrm{score}$ to include the proportion of candidates matching to radio sources (radio association).
 
%



In the fourth step, we applied all criteria to 234261 IRAS point sources, identifying \HII\ region candidates. After checking the quality of fluxes, we filter the IRAS point sources with the inequality of each criterion (see Table~\ref{tab:criteria}), yielding the \HII\ region candidates. In order to check the quality of criteria, we matched the \HII\ region candidates to radio continuum emission within a radius of  1$'$ (following~\citet{1997MNRAS.291..261W}), and the proportions (radio association) are listed in Table~\ref{tab:criteria}.

Because the radio association is also an important indicator of the quality of  the criteria, we modified the $F_1$ score and adopt a new type of score which is
 \begin{equation}
  \label{eq:score}
\mathrm{score}=\frac{3}{\left(1/\mathrm{sensitivity}+1/\mathrm{precision} + 1/\mathrm{radio\, association}\right)},
\end{equation}
where score is the harmonic mean of sensitivity, precision, and radio association.

In Table~\ref{tab:criteria}, we list the parameters of all criteria, and from left to right, the columns are the identity, the criterion, the number of selected candidates, the proportion of candidates associated with radio continuum (radio association), sensitivity, precision, the number of IRAS sources tested by the inequality of the criteria, and the score. We sort the criteria according to their scores.

\begin{figure*}
 
 \subfloat[Criterion 1.]{  \includegraphics[width = 0.45\textwidth]{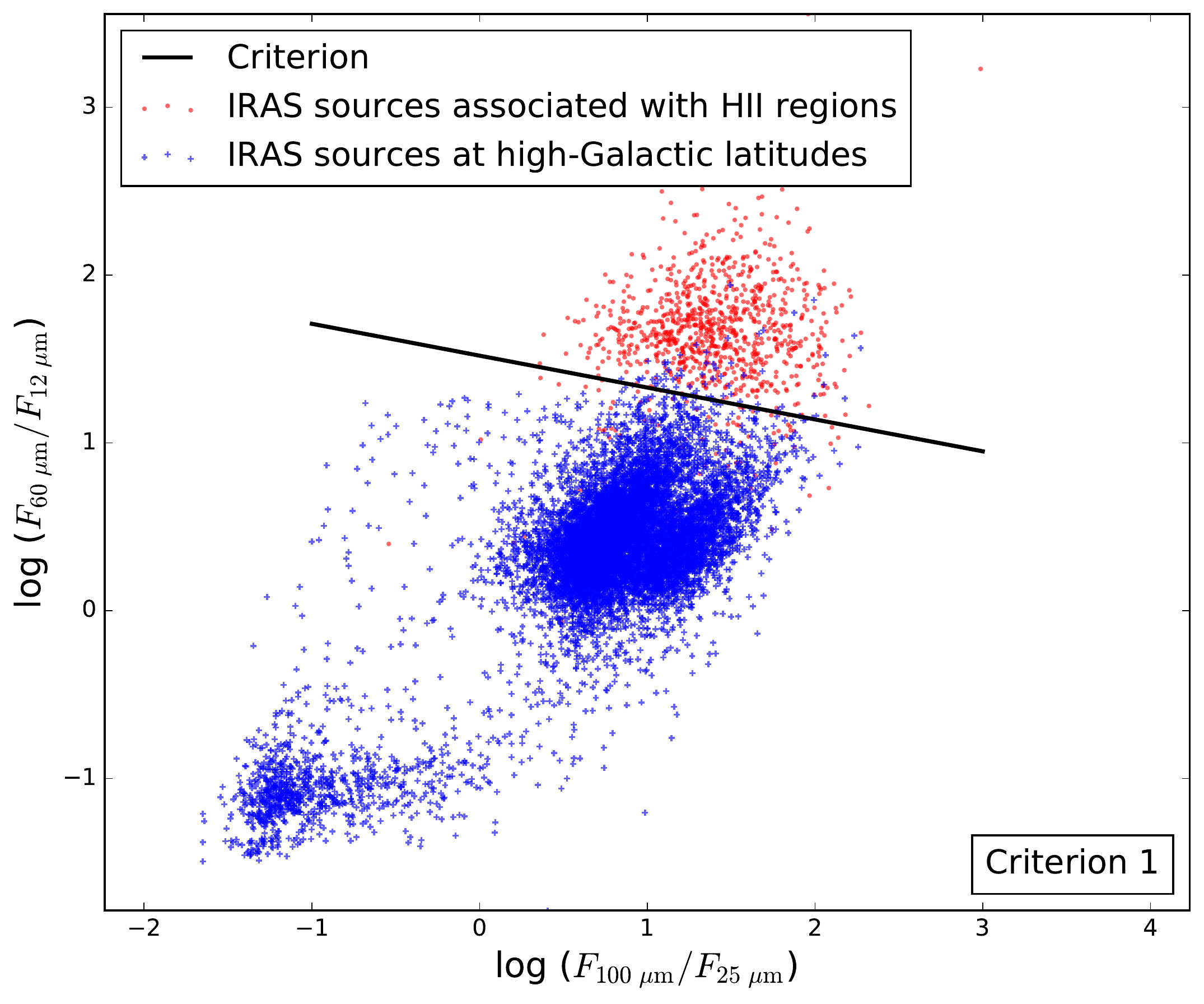}\label{fig:criterion1}} 
 \subfloat[Criterion 2.]{  \includegraphics[width = 0.45\textwidth]{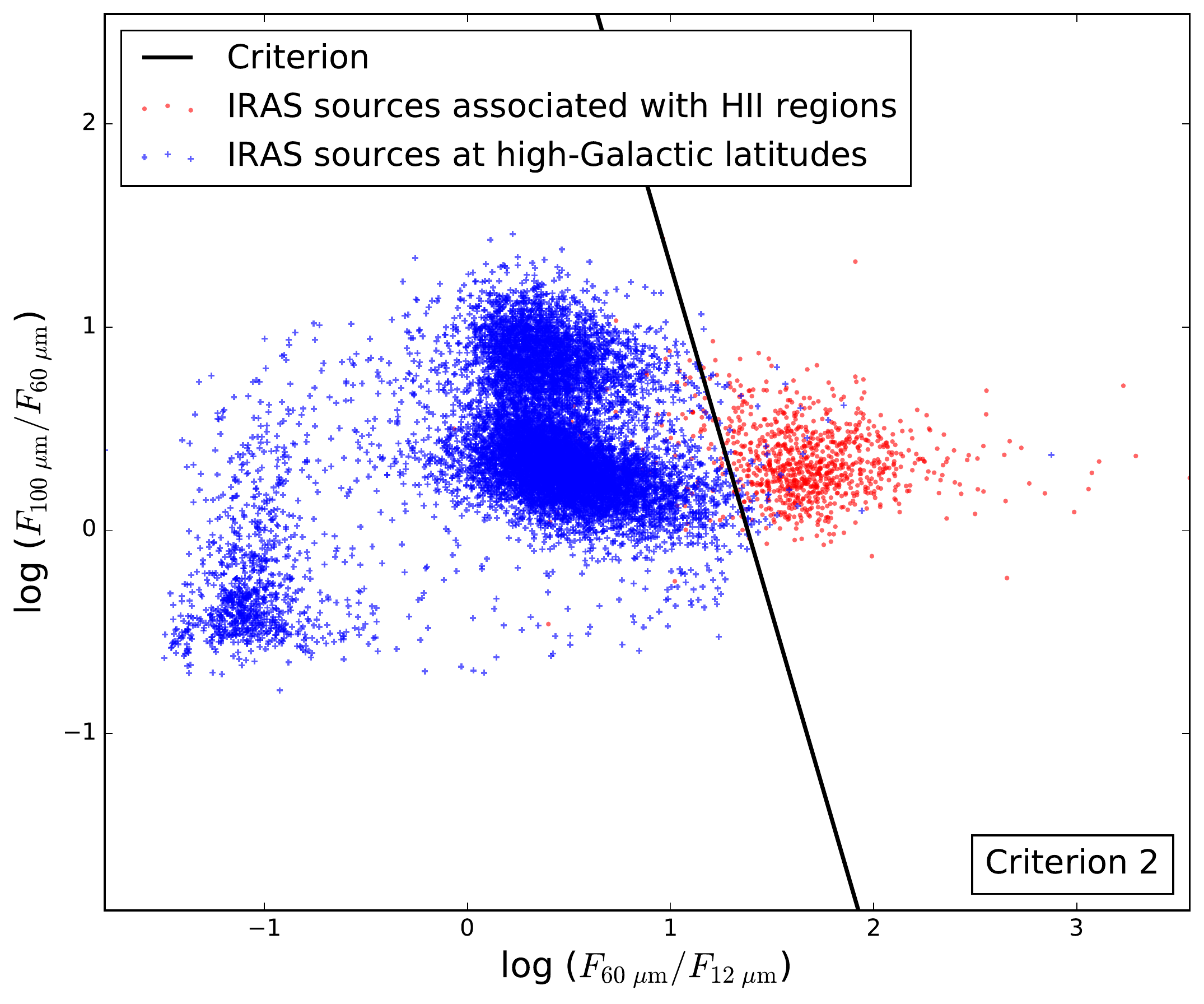}} \\
 \subfloat[Criterion 3.]{  \includegraphics[width = 0.45\textwidth]{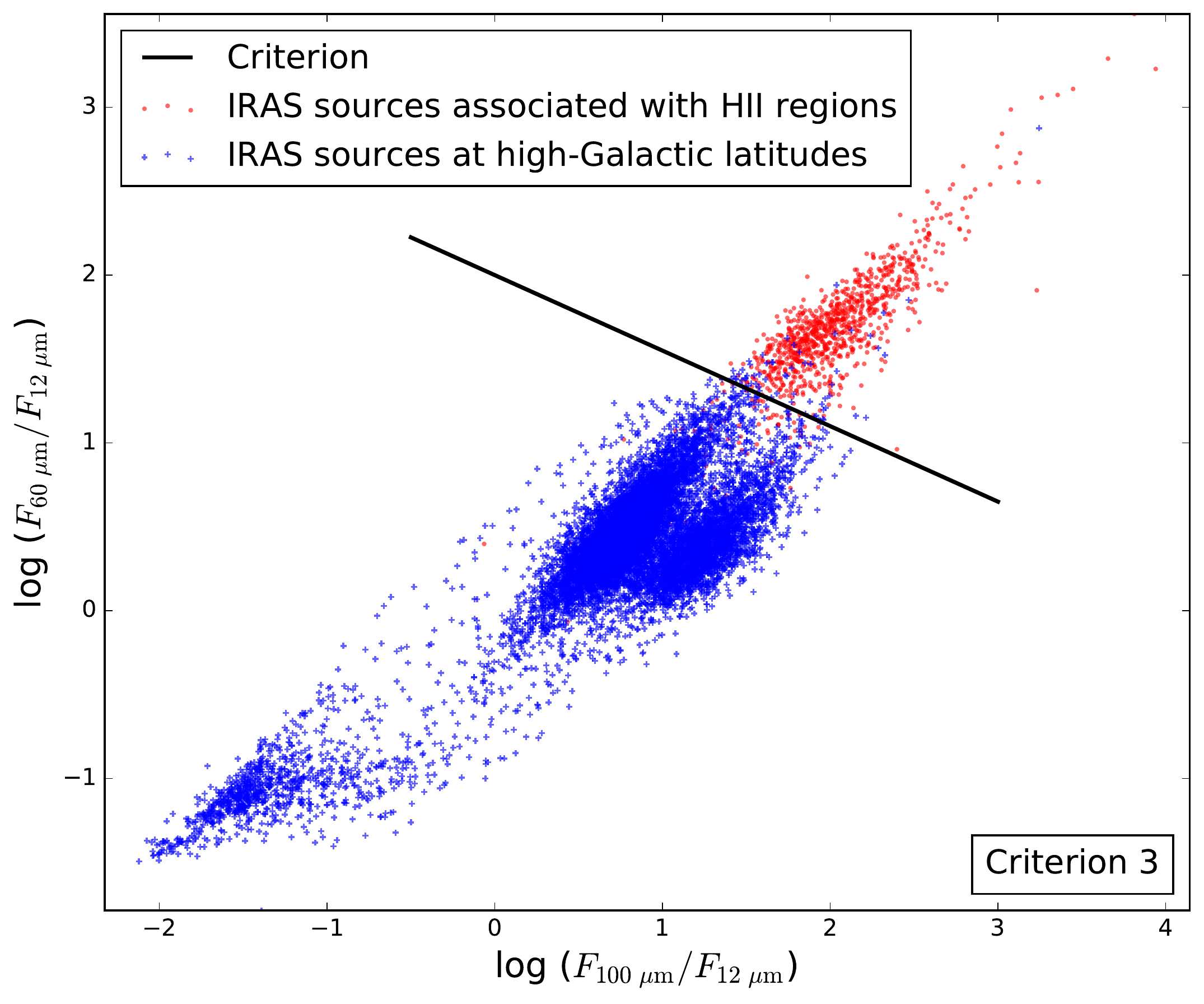}} 
 \subfloat[Criterion 4.]{  \includegraphics[width = 0.45\textwidth]{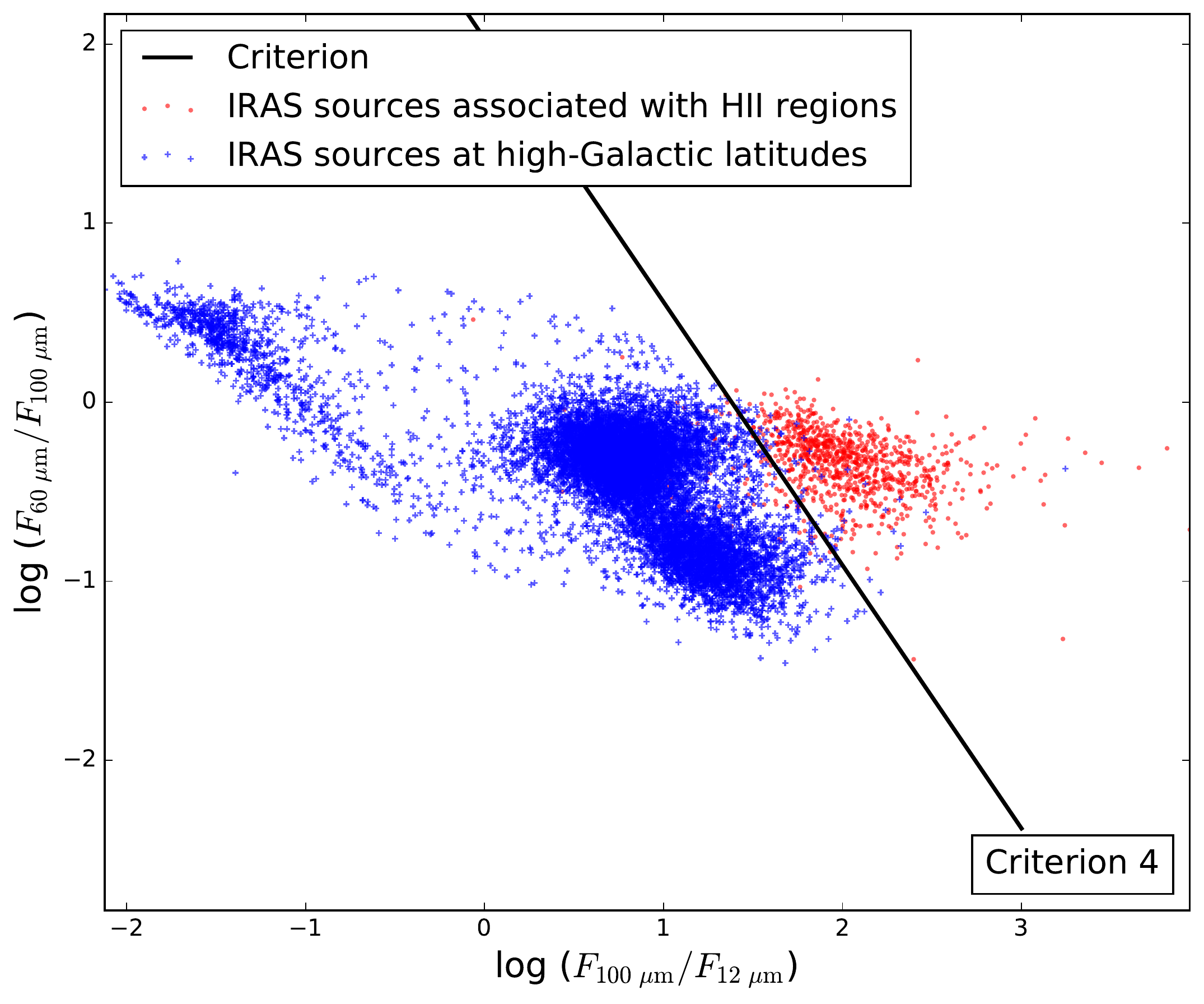}} \\
 \subfloat[Criterion 5.]{  \includegraphics[width = 0.45\textwidth]{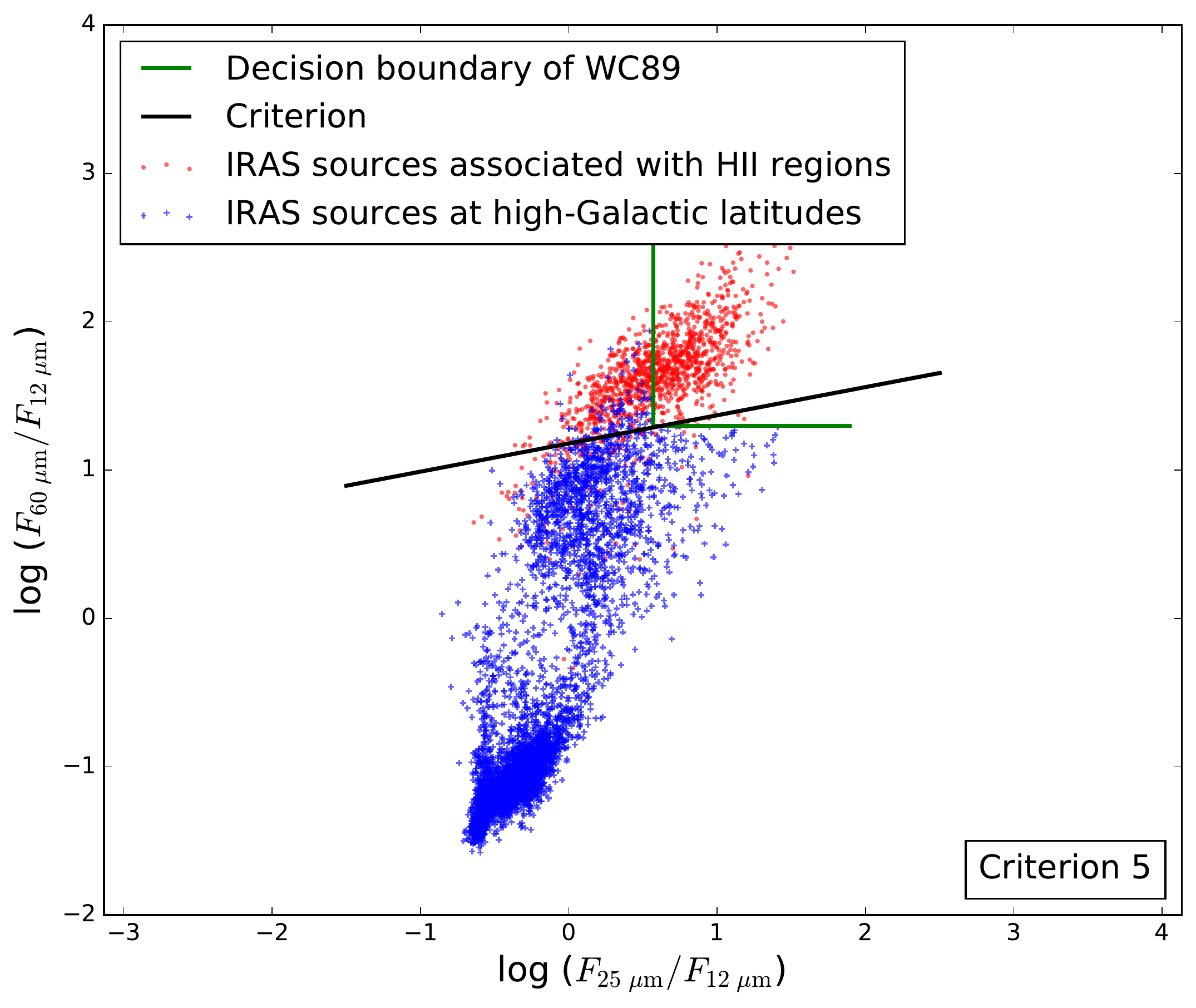}\label{fig:criterion5}} 
 \subfloat[Criterion 6.]{  \includegraphics[width = 0.45\textwidth]{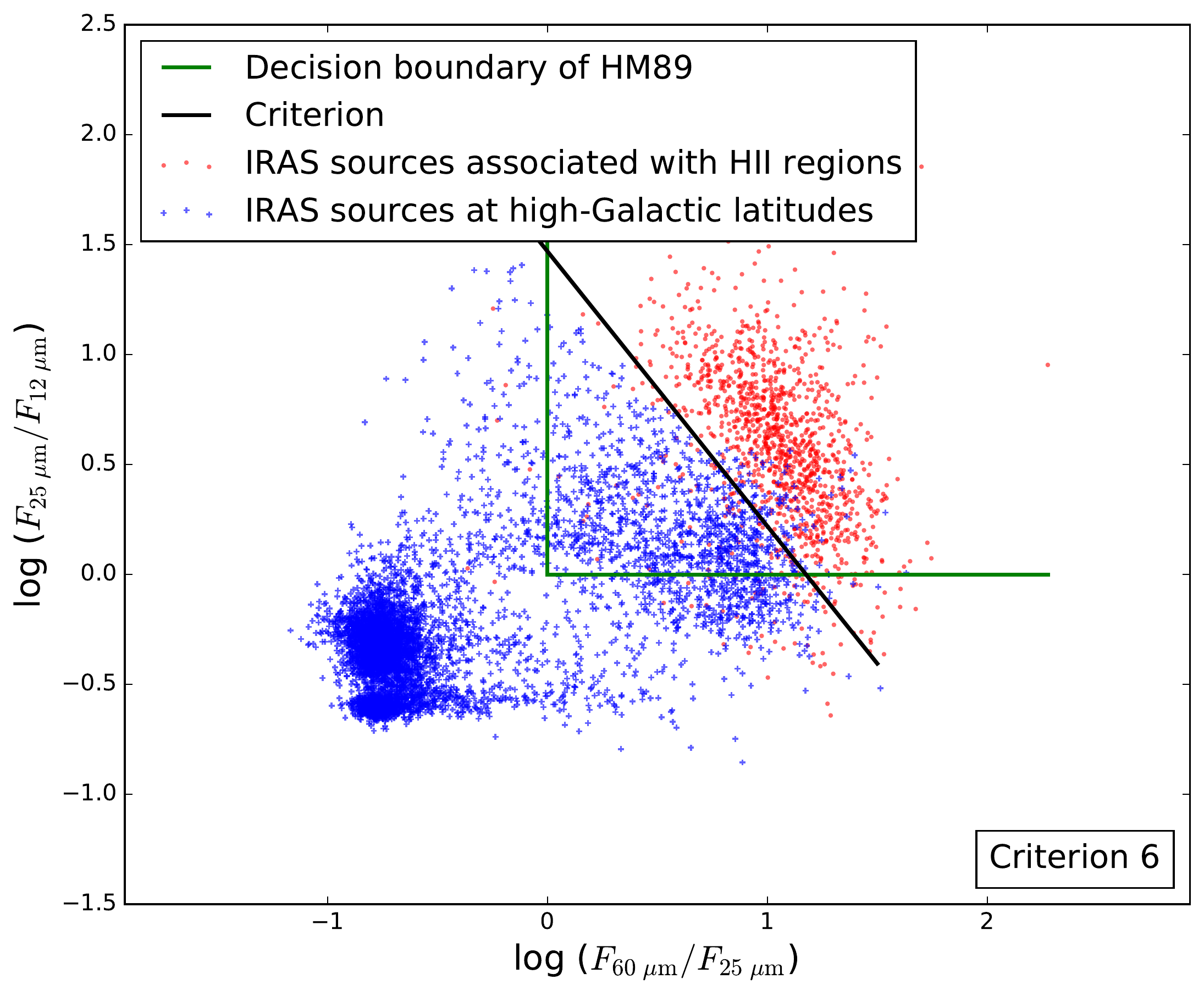}\label{fig:criterion6}} \\
  
\caption{ Top six criteria of Table~\ref{tab:criteria}. The criteria are criterion (a) 1,  (b) 2, (c) 3, (d) 4, (e) 5, and (f) 6. The red and blue markers represent \HII\ regions and non-\HII\ regions, respectively. The black lines denote the decision boundary (given by SVMs) of two groups of points.  The green lines in criterion 5 \& 6 delineate the decision boundary of  WC89 and HM89. }
\label{fig:compare}
\end{figure*}
\begin{figure*}
 
 \subfloat[Criterion 1.]{  \includegraphics[width = 0.45\textwidth]{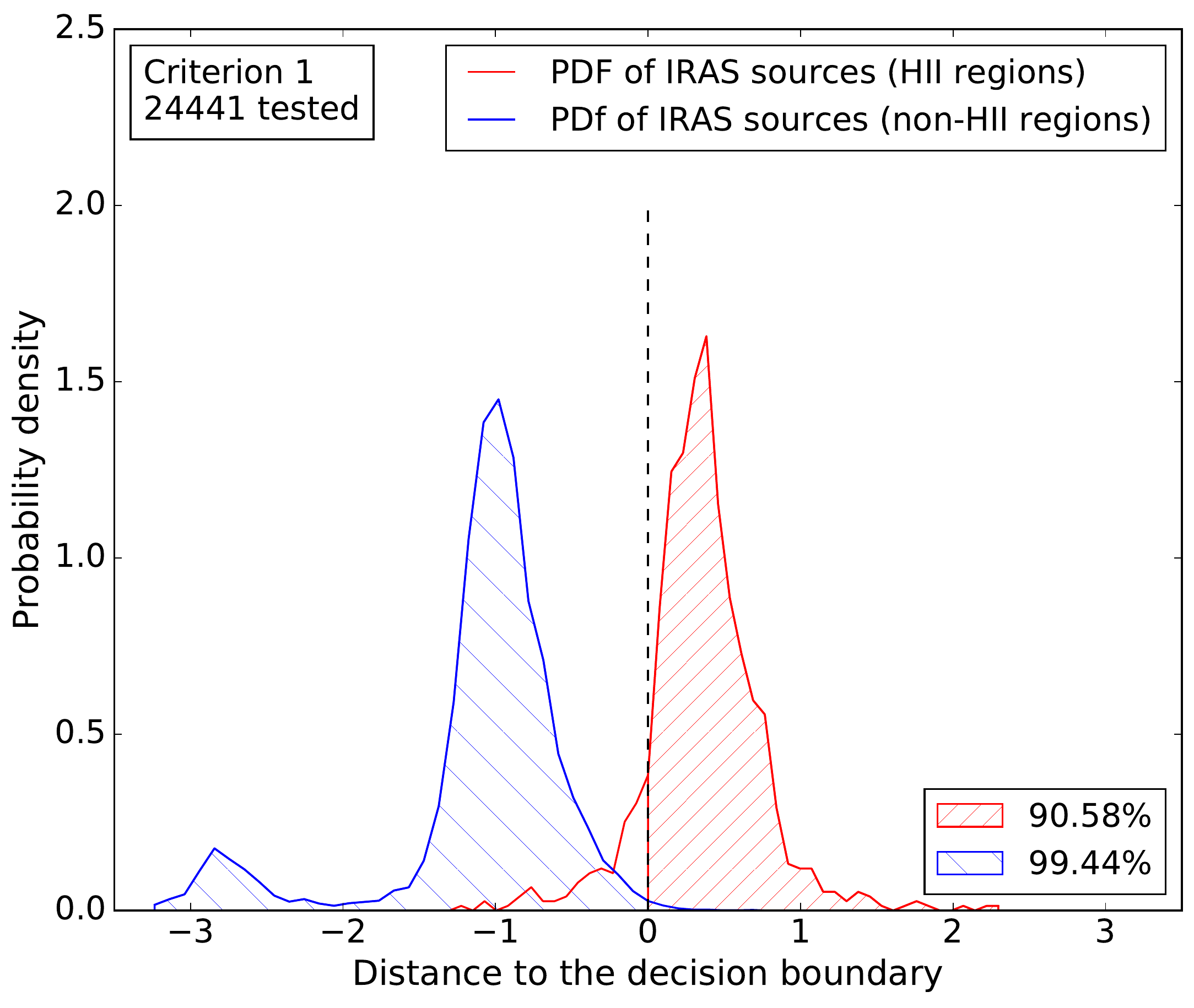}} 
 \subfloat[Criterion 2.]{  \includegraphics[width = 0.45\textwidth]{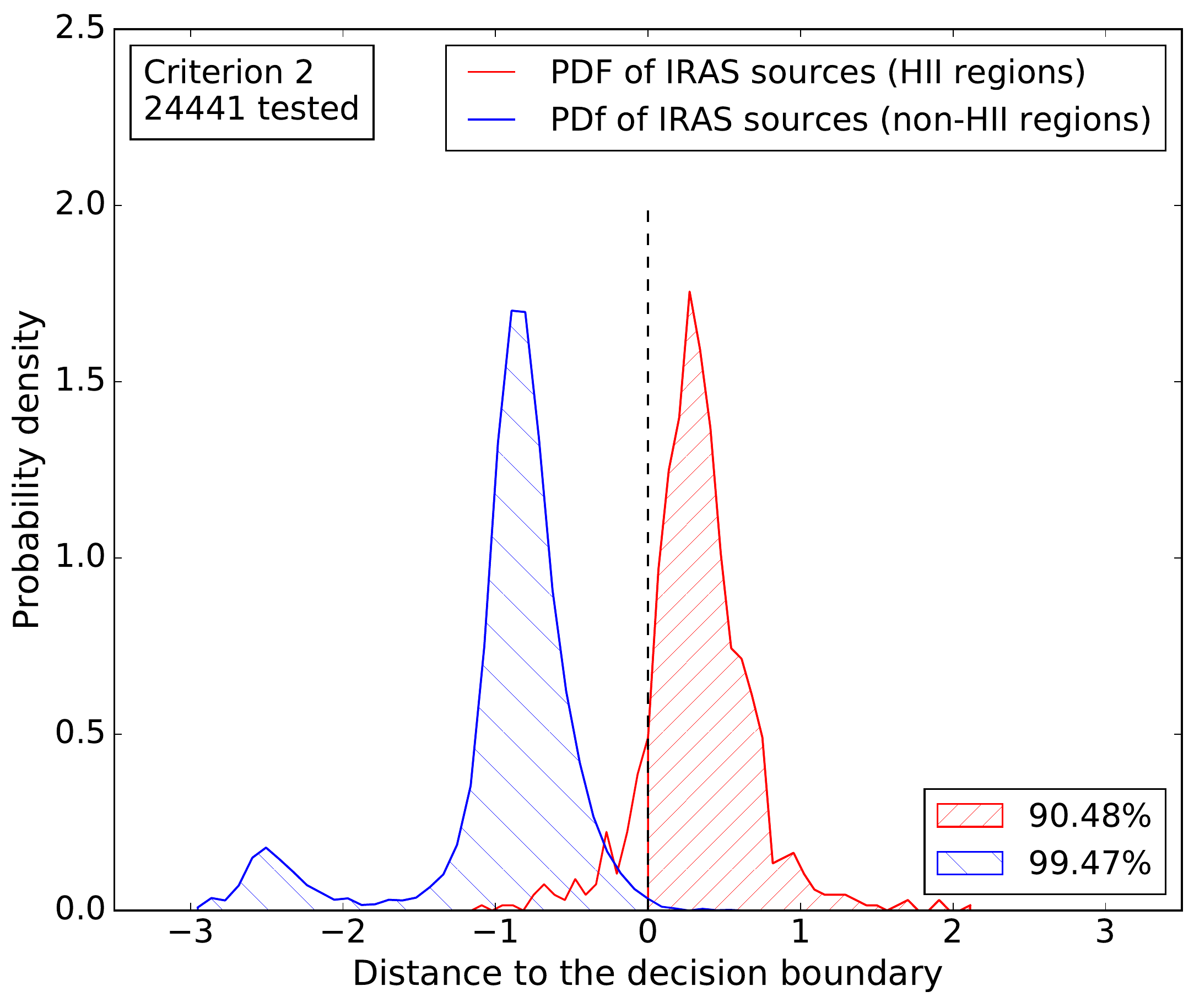}} \\
 \subfloat[Criterion 3.]{  \includegraphics[width = 0.45\textwidth]{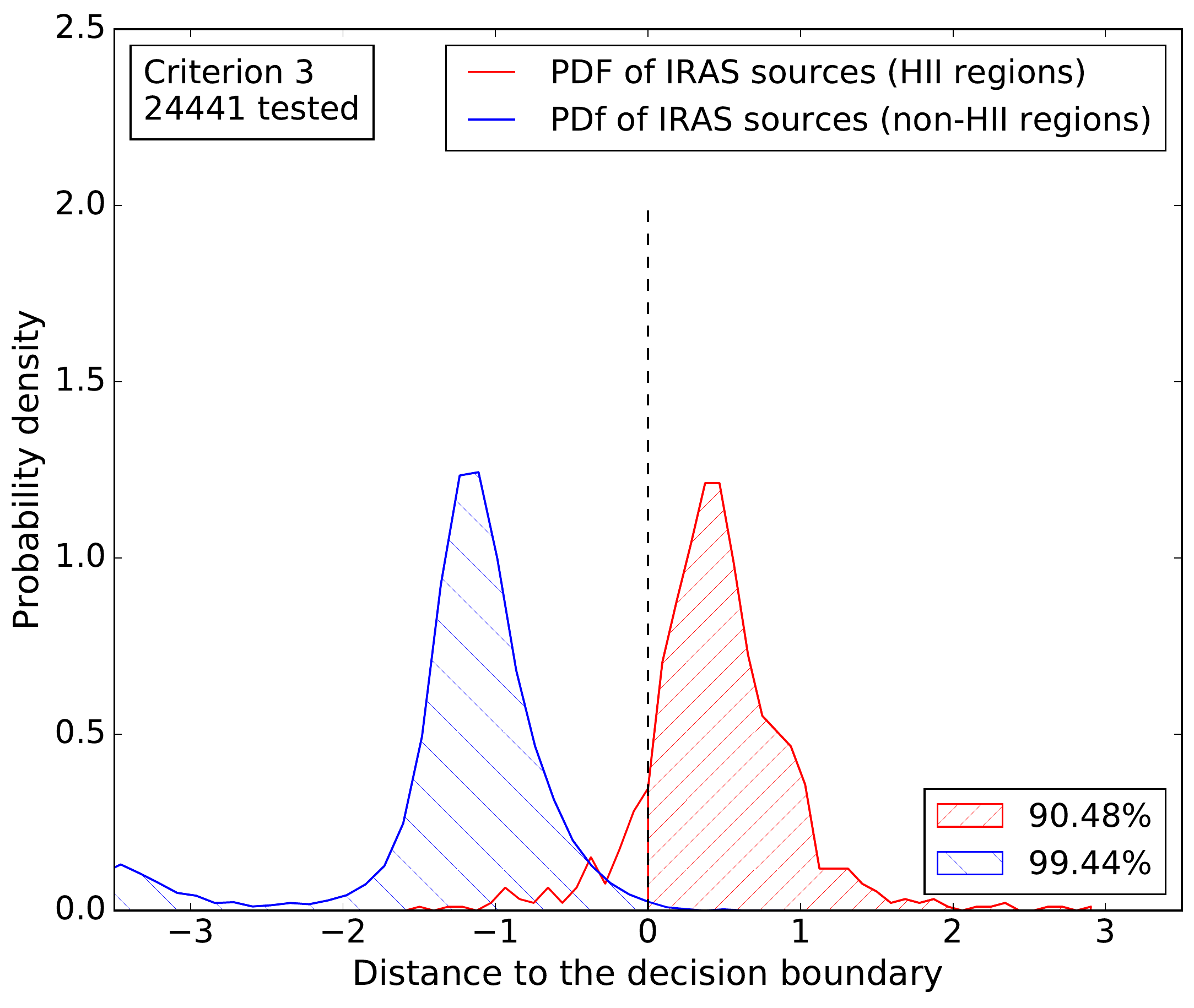}} 
 \subfloat[Criterion 4.]{  \includegraphics[width = 0.45\textwidth]{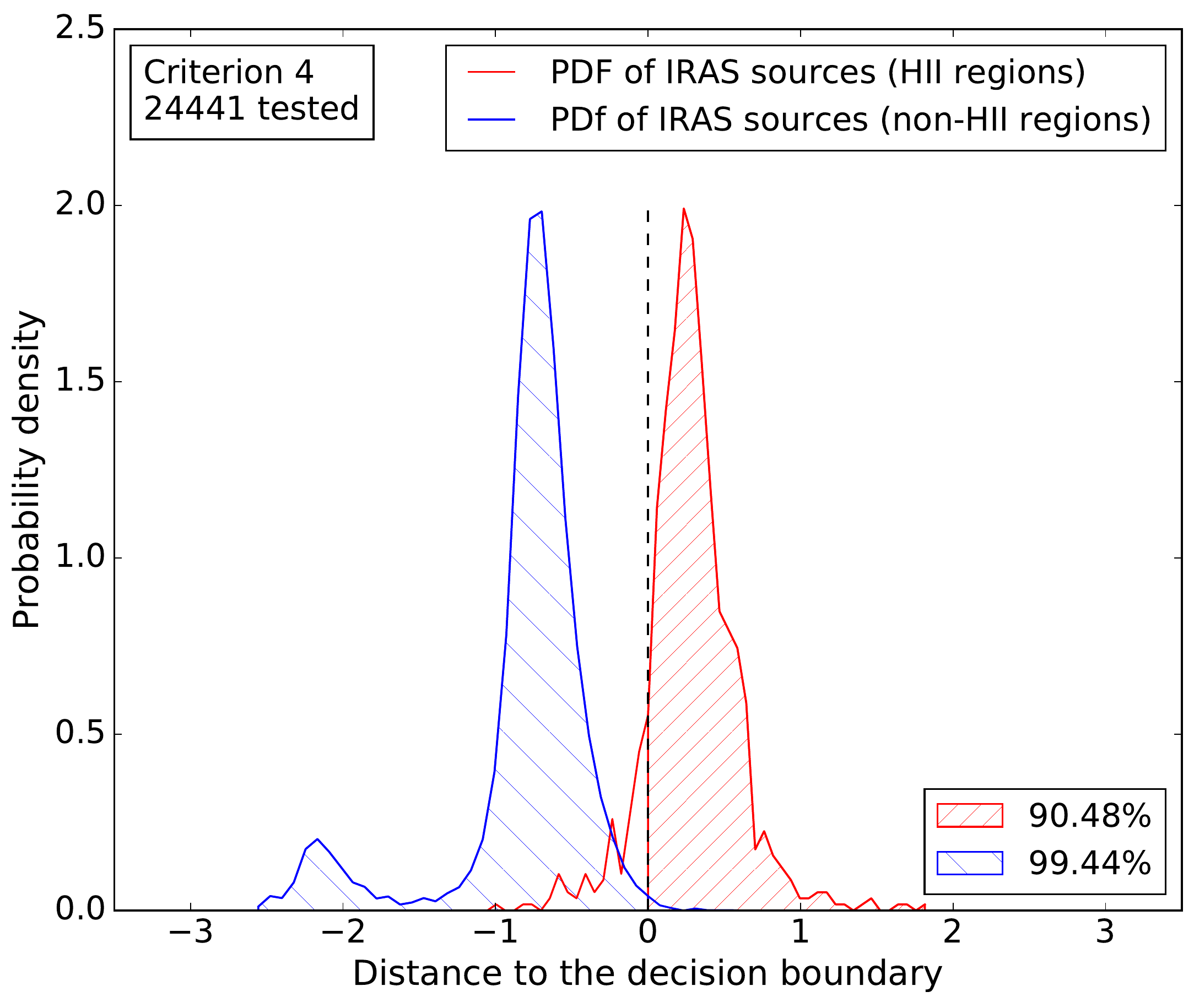}} \\
 \subfloat[Criterion 5.]{  \includegraphics[width = 0.45\textwidth]{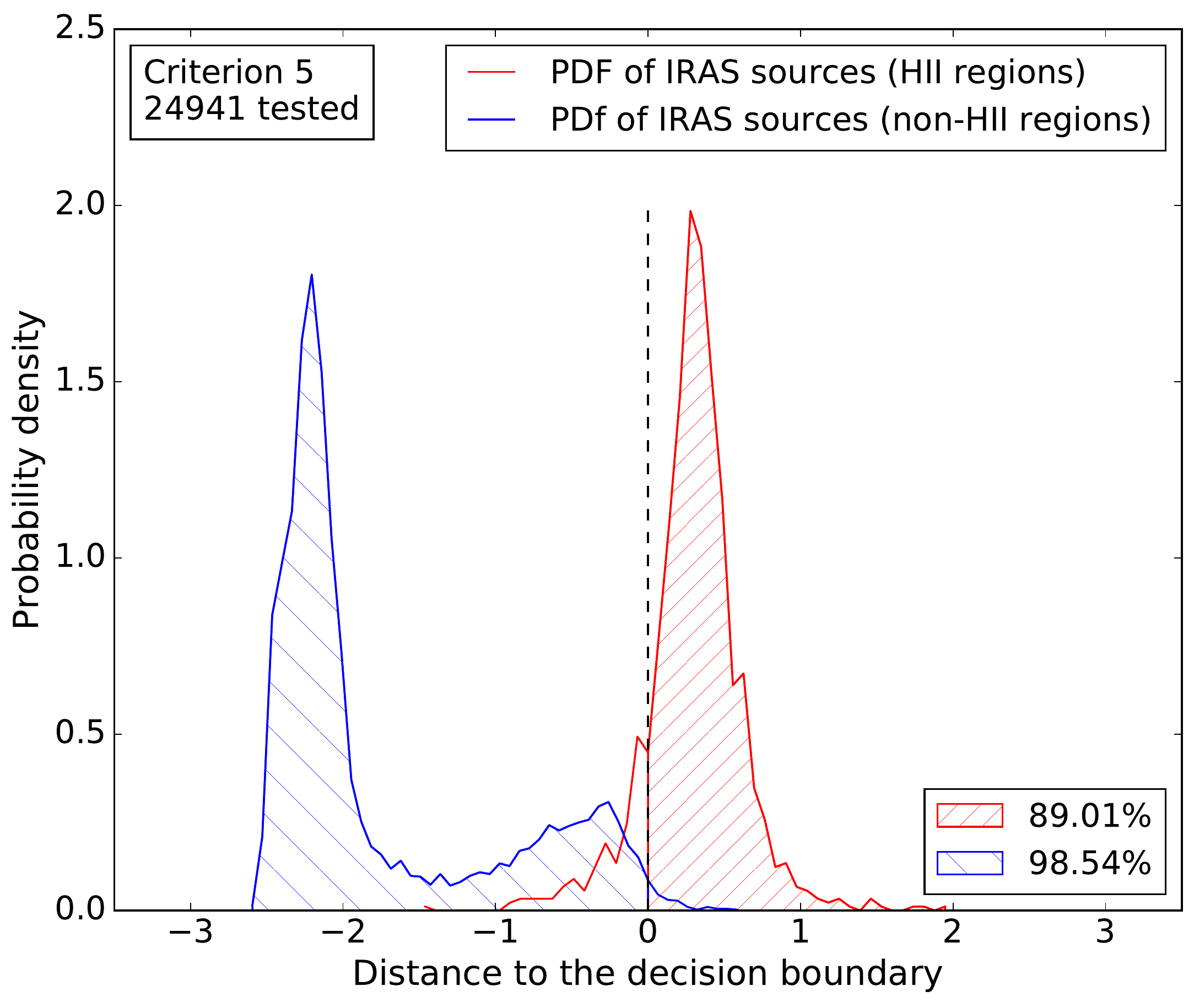}\label{fig:pdf5}} 
 \subfloat[Criterion 6.]{  \includegraphics[width = 0.45\textwidth]{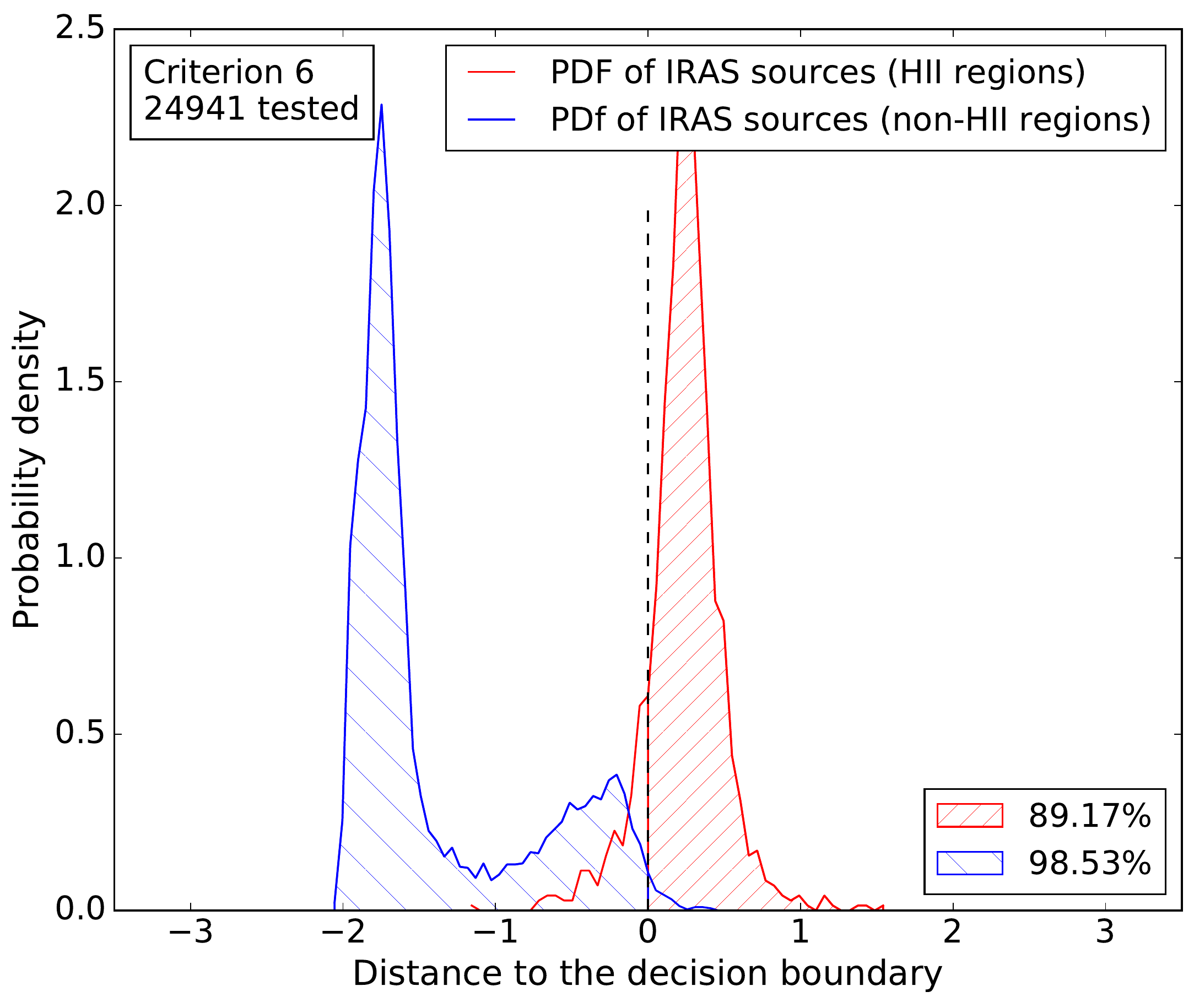}\label{fig:pdf6}} \\
  
\caption{PDFs of the top six criteria in Table~\ref{tab:criteria}. The red (\HII\ regions) and blue (non-\HII\ regions) lines represent the probability density of  IRAS sources with respect to the distance to the decision boundaries. Negative distances mean that IRAS sources disagree with the criteria. The red and blue shadowed areas represent the sensitivity (see Table~\ref{tab:criteria}) and specificity ($TN/(TN+FP)$) of criteria, respectively. }
\label{fig:pdf}
\end{figure*}

 We display details of the top six criteria of Table~\ref{tab:criteria} in Figure~\ref{fig:compare}, where the criteria provided by HM89 and WC89 are delineated by green lines. Figure~\ref{fig:pdf} shows the distribution of IRAS sources in terms of distances to the decision boundary, with negative values signifying sources rejected by criteria.

In the final step, we determine the optimal criterion. Because criterion 1 possesses the highest score, we adopt criterion 1 as the optimal criterion.  In Table~\ref{tab:criteria}, criterion 2, 3, and 4 require  $Q_{60}>1$ and $Q_{100}>1$  and they all use 12, 60, and 100 \um\ bands, which means they share the same data. The resemblance of results between criterion 2, 3, and 4 indicates the robustness of SVMs, and as expected, criterion 1 shows slightly better results because it uses all four bands.  


 The sensitivity of criterion 1 indicates that $\sim$91\% of \HII\ regions have been identified and its specificity (see Figure~\ref{fig:pdf}) shows that less than 1\% of high Galactic latitude sources are \HII\ regions. The radio association of criterion 1  is approximately 2\% higher than the proportion (32\%) in \wise\ \HII\ region catalogue, which gives us a high confidence in the reliability of \HII\ region candidates.

 \subsection{SVMs versus LDA}


In this subsection, we compare the decision boundary determined by SVMs to that produced by linear discriminant analysis (LDA), which is another method to decide a linear decision boundary.  In two-dimensional cases, LDA attempts to determine a decision boundary by maximizing the separation between samples and minimizing their scatter for the  projections on an axis perpendicular to the decision boundary. However, SVMs primarily focus on the points near the boundary of samples.
 
\begin{figure}
 
\includegraphics[width = 0.45\textwidth]{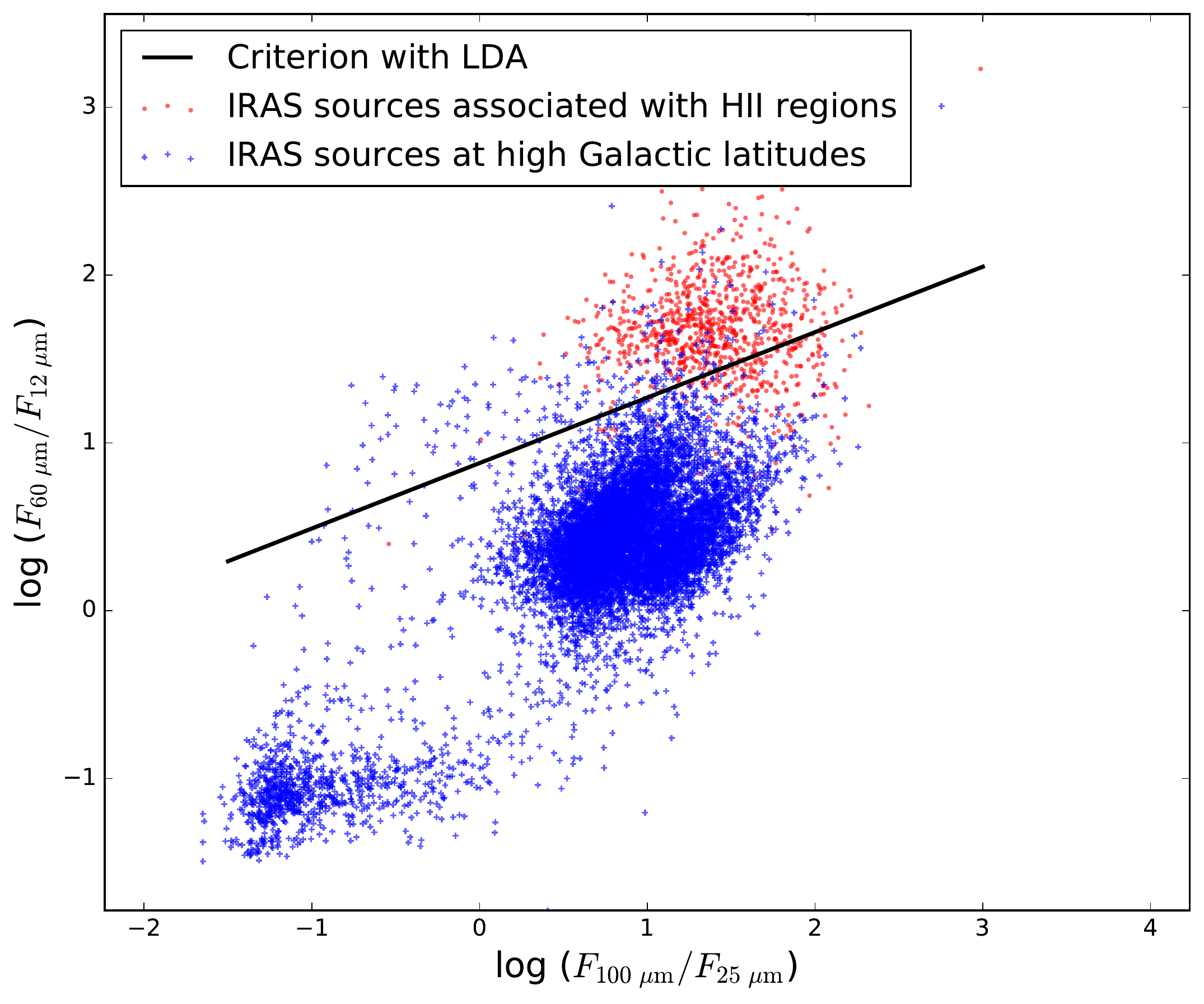}

\caption{ Decision boundary determined by LDA.  The red and blue markers represent \HII\ regions and non-\HII\ regions, respectively.  }
\label{fig:lda}
\end{figure}

 We performed LDA with the \emph{sklearn} package with default parameters. As an example,  we display the results of colour combination of criterion 1. In Figure~\ref{fig:lda}, we illustrate the decision boundary determined by LDA.  Evidently, LDA misclassifies many more \HII\ regions and includes  many more  non-\HII\ regions that the SVM algorithm (see Figure~\ref{fig:criterion1}).


 Consequently, the decision boundary determined by SVMs gives better results than that determined by LDA for selecting \HII\ regions from the IRAS PSC.  
 

\subsection{Three-colour criteria}
 

Because the IRAS PSC provides fluxes for four bands, we also examined  three-colour criteria. The process of creating three-colour criteria is identical to that of two-colour criteria. We also require that the flux quality of the numerator is better than an upper limit (above the detection threshold). 

 The most important factor that  affects the results of three-colour criteria is which bands are required to be better than an upper limit ($>1$).  In each three-colour criterion, we already used all four bands, and  the rearrangement of the colour combinations of these four bands does not lose or gain information. Consequently, if two three-colour criterion requires the qualities of  same bands  to be better than an upper limit ($>1$), they essentially give the same results, and two-colour criteria (which involve three bands) shows similar results, for instance, the resemblance of criterion 2, 3, and 4 in Table~\ref{tab:criteria}.





 We derived three-colour criteria for all 16 possible three-colour combinations. The three-colour criteria possessing the highest score (0.565) is given by 
\begin{equation}
   \left\{
	\begin{array}{c}
\label{eq:threeCriteria}
4.93 \times F_{60}/F_{12}+0.17\times F_{60}/F_{25}+1.42\times F_{100}/F_{60}> 6.95,\\
Q_{60}>1, \\Q_{100}>1,
\end{array}
\right.
\end{equation}  
which identifies 3093 candidates. The radio association, sensitivity, and precision are 31.88\%, 90.48\%, and 93.70\%, respectively. Those parameters are close to that of criterion 2. However, the score is lower than criterion 1. Intuitively, three-colour criteria should do better than two-colour criteria, but the four bands are not entirely independent and the decision boundary in three-dimensional space may not be well constrained.   Consequently, we only consider two-colour criteria for selecting \HII\ regions from the IRAS PSC.

\section{Results}
\label{sec:result}
 

In this section, we examine the distribution of \HII\ candidates selected from IRAS PSC,   the total number of \HII\ regions in the Milky Way, and the evolution of \HII\ regions on the two-colour diagram,

    \begin{figure*}
 \centering
 \includegraphics[width = 0.8\textwidth]{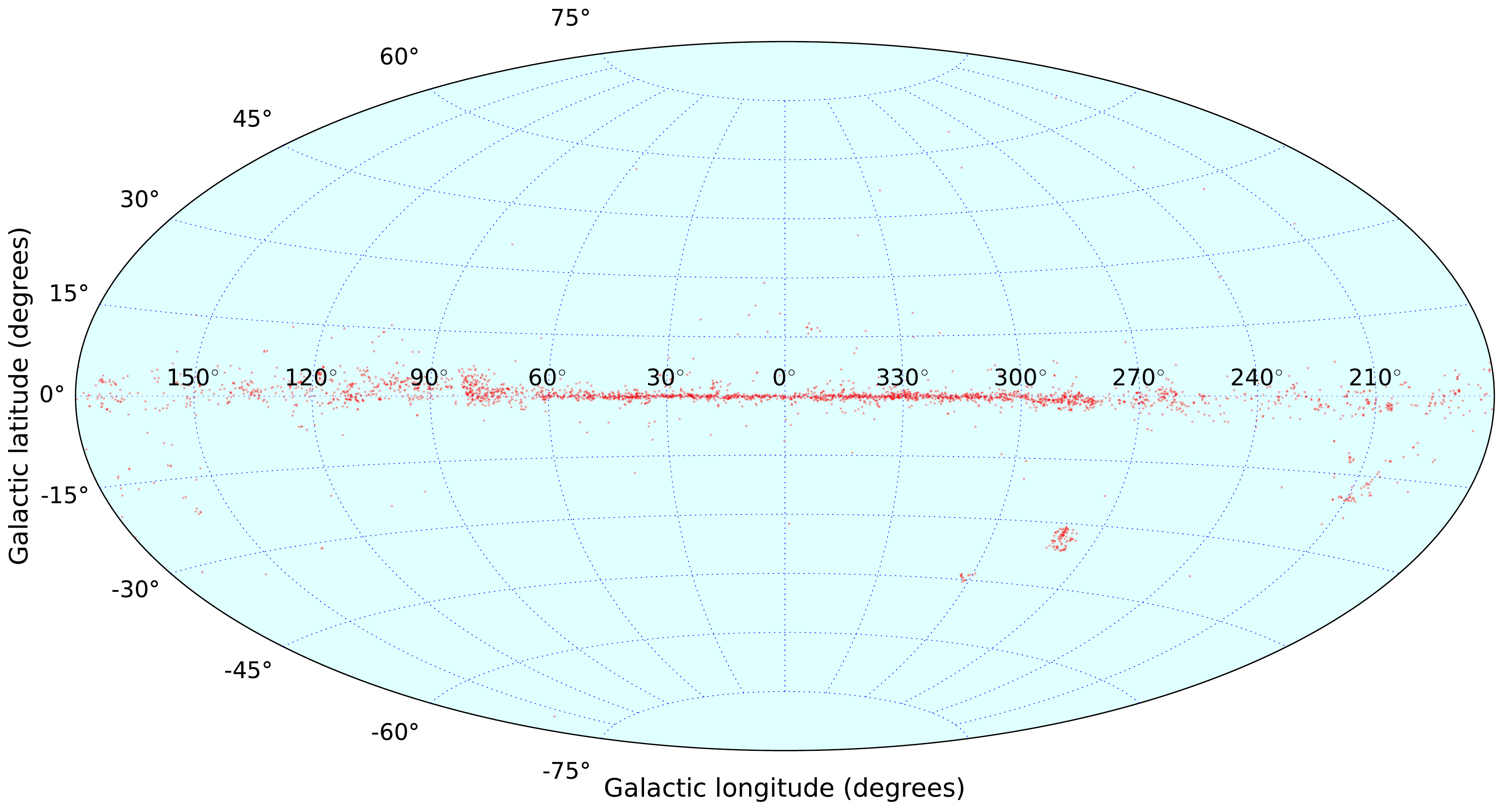}  
 
\caption{All-sky distribution of  3041  IRAS \HII\ region candidates  selected by criterion 1. After eliminating  sources  in LMC and SMC and that have been identified as PNe, Post-AGB stars, supernova remnants,  and extragalactic objects via SIMBAD catalogues,  2805 \HII\ region candidates remain.}
\label{fig:allsky}
\end{figure*}

     \begin{figure*}
   \subfloat[The histogram along the Galactic latitude.]{ \includegraphics[width = 0.46\textwidth]{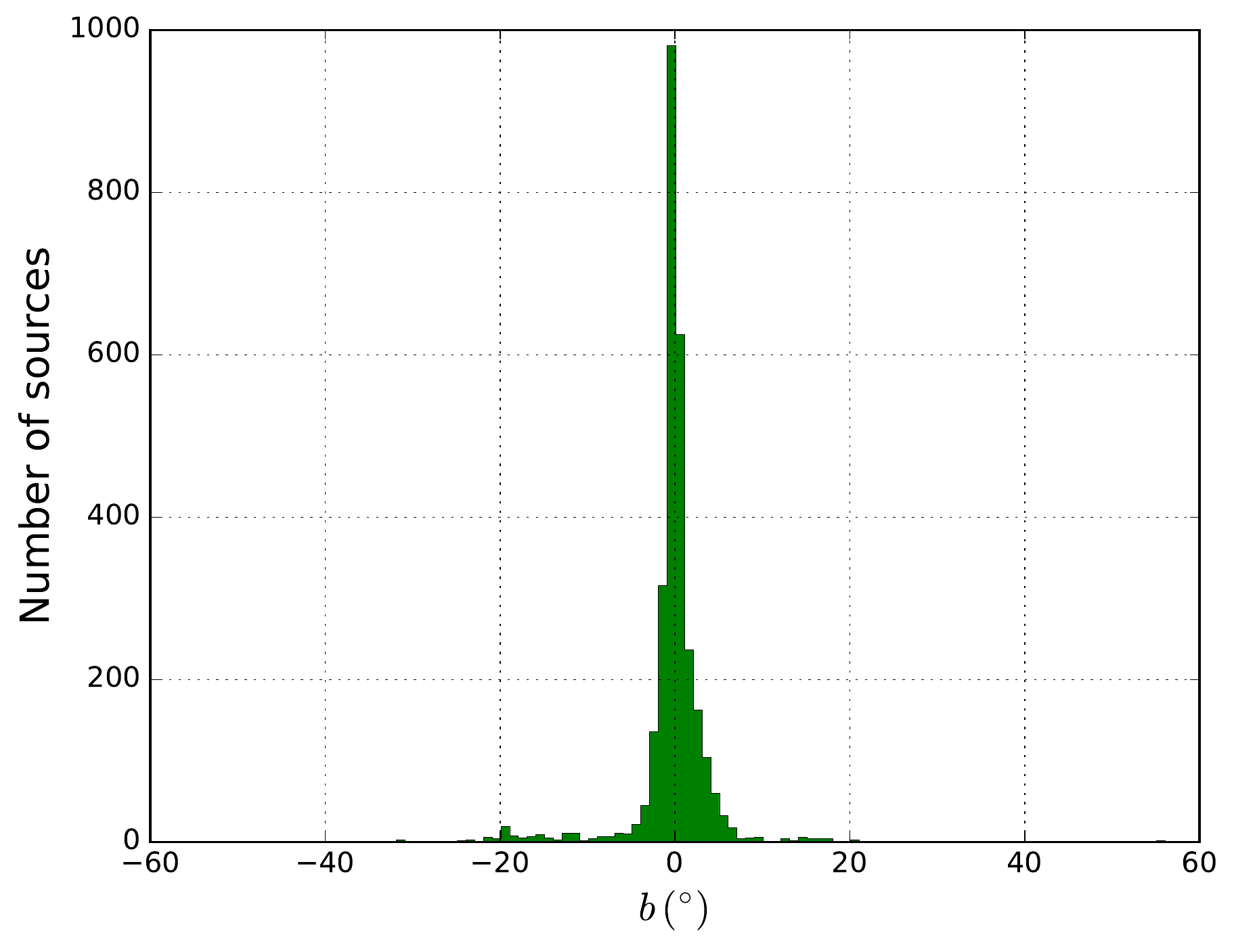}} 
  \subfloat[The histogram along the Galactic longitude.]{  \includegraphics[width = 0.44\textwidth]{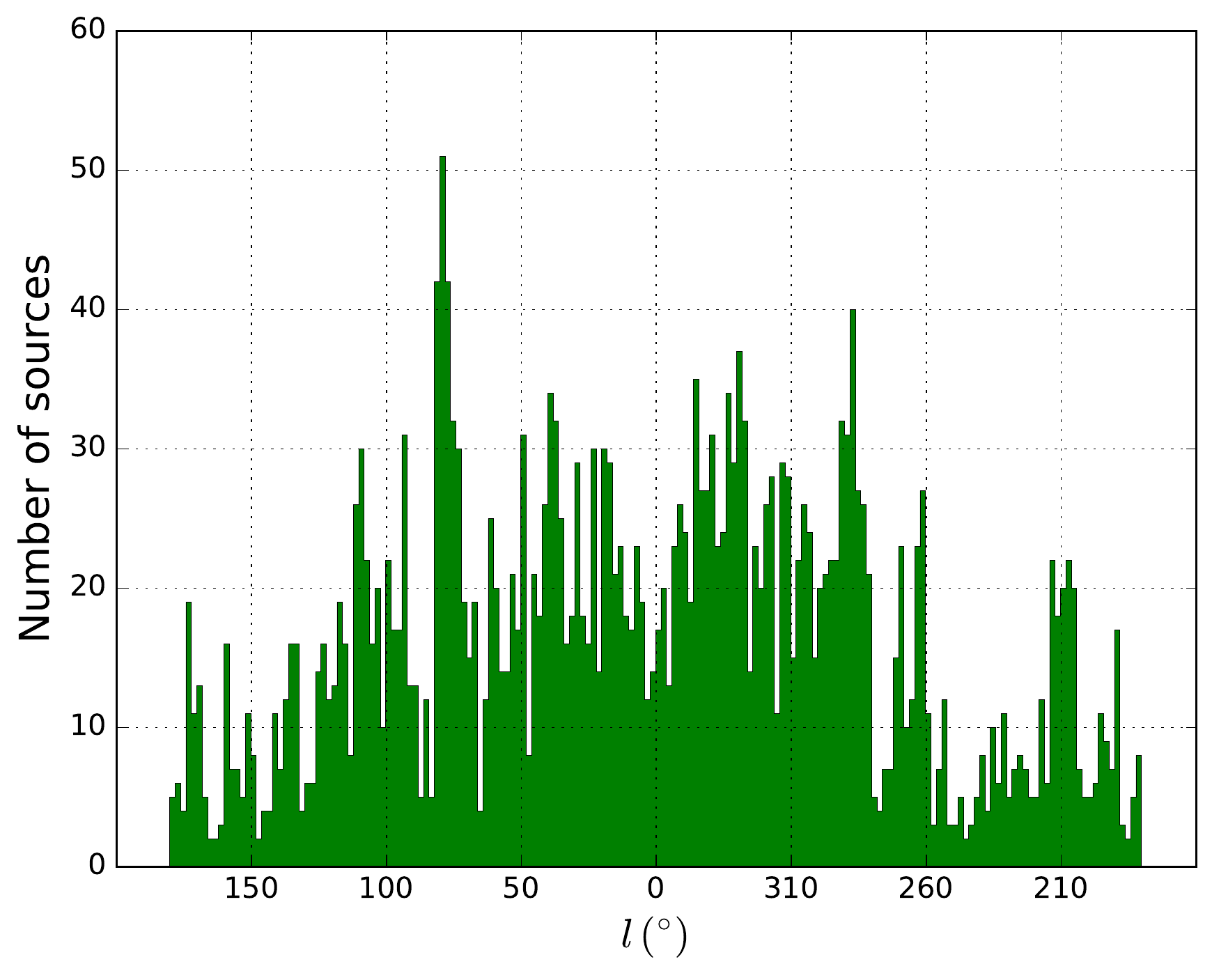}} \\
\caption{ Histograms of the distribution of \HII\ region candidates identified by criterion 1 along the Galactic latitude and longitude, the bin sizes of which are 1\degr\ and 2\degr, respectively.}
\label{fig:latitude}
\end{figure*}

 \subsection{Source distribution}

We display the all-sky distribution of those sources selected by criterion 1 in Figure~\ref{fig:allsky}. As illustrated in Figure~\ref{fig:allsky}, the Galactic plane is clearly delineated by those 3061 \HII\ region candidates,  and the sources at high Galactic latitudes are sparse.  As expected, the LMC and the SMC are evident, as well as the Orion nebula.  We examined 17 selected  high Galactic latitude sources ($|b|>30^\circ$, not in SMC or LMC) and found that 14 of them have been identified as galaxies or extragalactic \HII\ regions. The remaining three sources are one Planetary Nebula (PN), one Post-AGB star, and one unknown infrared source (IRAS 13458-0823).  

In Figure~\ref{fig:latitude}, we plot histograms of these sources along the Galactic latitude and Galactic longitude axes, the bin sizes of which are 1\degr\ and 2\degr, respectively. The distribution of \HII\ region candidates along the Galactic latitude is approximately Gaussian, despite  two prominent groups at high Galactic latitudes, associated with the Orion nebula and the LMC. In the right panel  of Figure~\ref{fig:latitude}, two peaks are evident near 80\degr\ and 280 \degr, corresponding to the Local Arm and  Carina Arm~\citep{1985SvAL...11..185A, 1989AJ.....97..786H, 2013ApJ...769...15X}, respectively.

 After eliminating the sources in LMC, SMC, we have 2940 \HII\ region candidates left. We further reject those IRAS sources that have been identified as PNe, Post-AGB stars, supernova remnants,  and extragalactic objects via SIMBAD catalogues, and 2805 ($=2940 -135$) \HII\ region candidates remain. 68\% of those remaining   (2805)  \HII\ candidates are in the first and fourth Galactic quadrants (within 90\deg of the Galactic Centre), and  47\% are within 60\deg of the Galactic Centre. Compared with the proportion of \wise\ catalogue ( 86\% and 76\%), those proportions are smaller. This is because the spatial resolution of the IRAS is not sufficiently good  to resolve many small-angular-size \HII\ regions in the first and fourth Galactic quadrants. This can be demonstrated by the fact that the proportions of   large (radii>1\arcmin) and small (radii<1\arcmin)   angular size  \wise\ \HII\ region candidates   in the  first and fourth Galactic quadrants   are 76\% and 98\%, respectively  The asymmetric distribution  of \HII\ regions in Galactic latitude is also present  in IRAS \HII\ candidates. 53\% of IRAS \HII\ candidates are at negative latitudes, and this is  close to the value of \wise\ catalogue (56\%).

 \subsection{The total number of \HII\ regions in the Milky Way}
\label{totalNumber}
 

      \begin{figure}
 
\includegraphics[width = 0.45\textwidth]{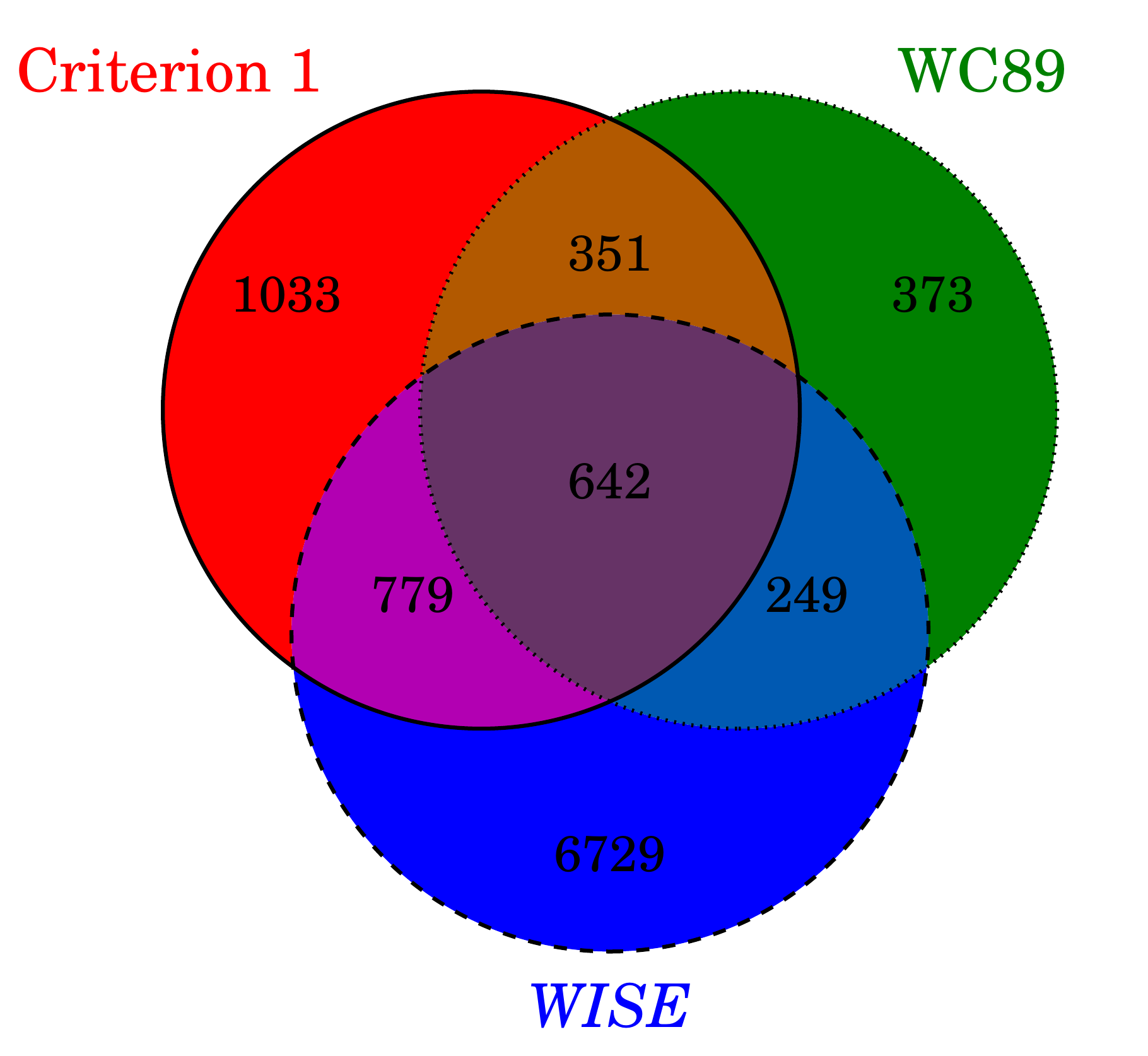}

\caption{ The Venn diagram of \HII\ region candidates in the Milky Way. The red, green, and blue colours represent IRAS sources selected by criterion 1, IRAS sources identified by  WC89~\citep{1989ApJ...340..265W},  and \wise\ \HII\ region candidates~\citep{2014ApJS..212....1A}, respectively. }
\label{fig:venn}
\end{figure}



We estimate the total number of \HII\ regions in the Milky Way by combining the \wise\ \HII\ region catalogue with the \HII\ region candidates in the IRAS PSC identified by criterion 1 or by WC89.

  In Figure~\ref{fig:venn}, we show a Venn diagram  with respect to criterion 1, WC89, and \wise\ \HII\ region candidate catalogues. Of the   2805  \HII\ region candidates selected according to criterion 1, 1421 candidates are positionally associated with the \wise\ \HII\ region catalogue (within 1\arcmin).   After eliminating non-\HII\ regions, WC89 identifies 1615 \HII\ region candidates,  and 993 of them agree with criterion 1.  As shown in Figure~\ref{fig:venn},  the total number of \HII\ regions   in the Milky Way is approximately  10156.


The number of $\sim$10200 is a lower limit of \HII\ regions in the Milky Way.  On one hand, the sensitivity of criterion 1 suggests  $\sim$91\% of \HII\ regions in the IRAS sources have been identified, missing a small   amount  of \HII\ regions; on the other hand, due to the lower spatial resolution and sensitivity, the IRAS catalogue cannot resolve many \HII\ regions whose angular sizes are small and may  have missed some \HII\ regions whose  infrared   emissions are beyond the detectability of IRAS.

      \begin{figure*}
 
  \subfloat[Criterion 1.]{  \includegraphics[width = 0.45\textwidth]{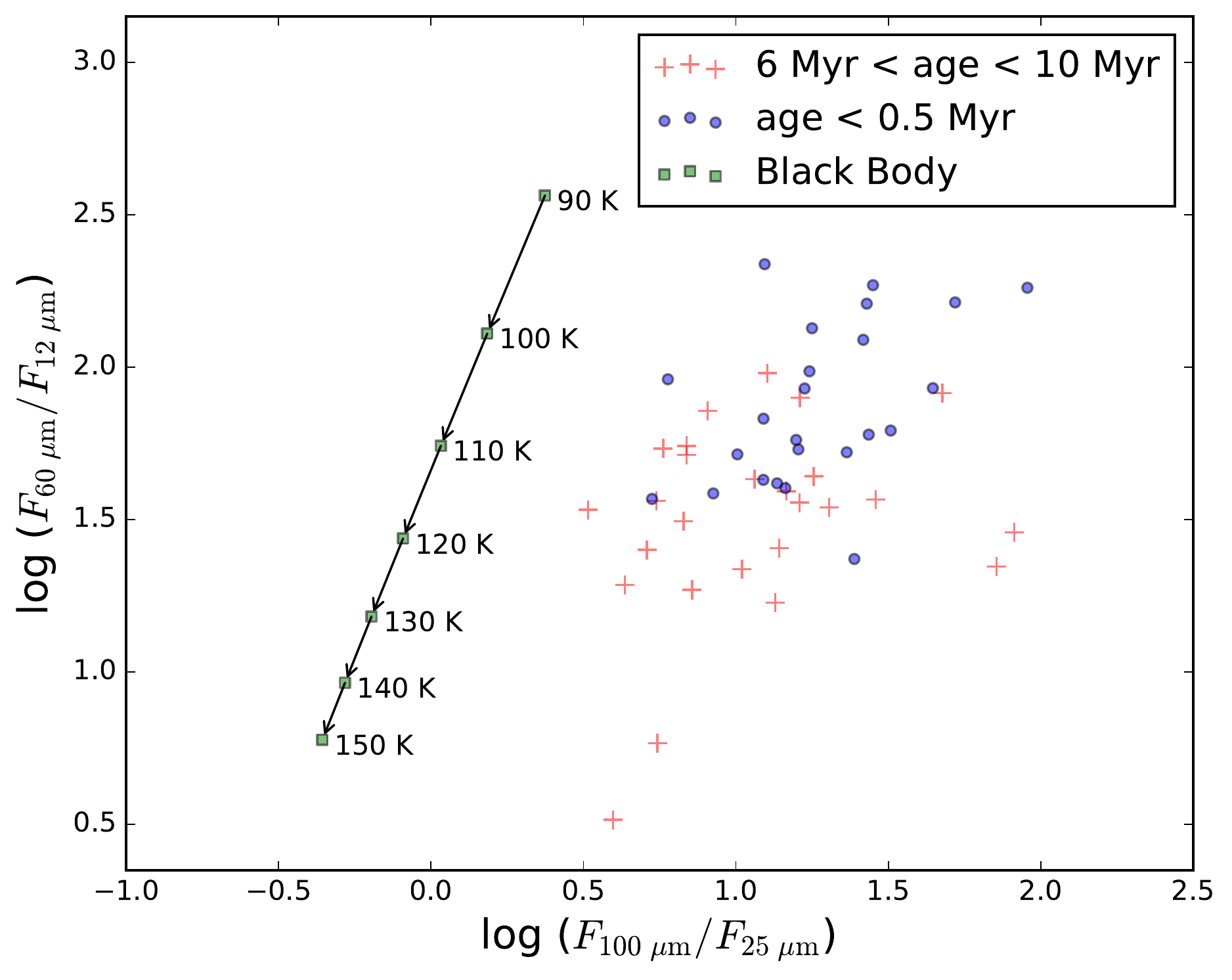}\label{fig:evocriterion1}} 
 \subfloat[WC89 (criterion 5).]{  \includegraphics[width = 0.45\textwidth]{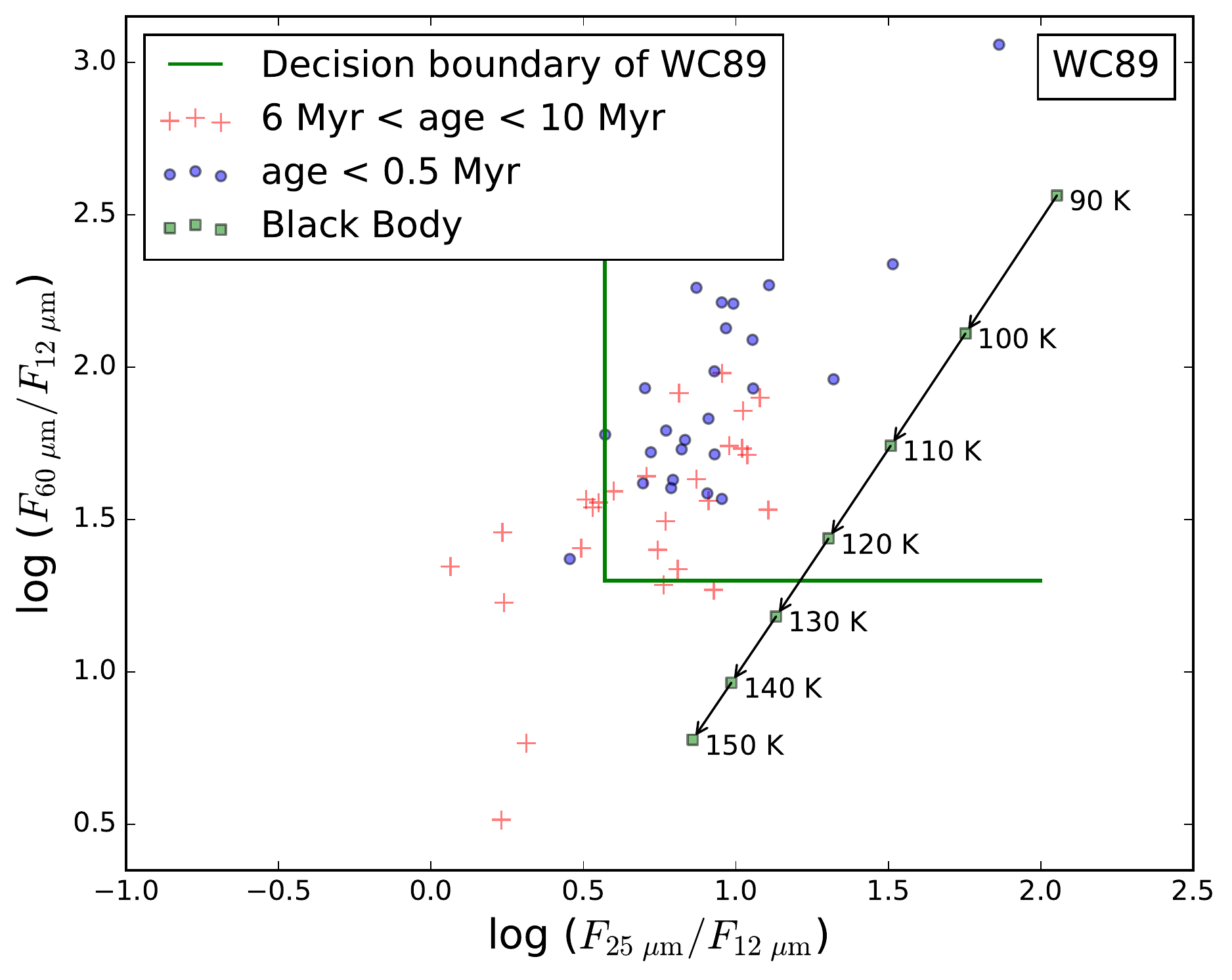}\label{fig:evowc89}} \\
\caption{The evolution of \HII\ regions on two-colour diagrams.  The colour combinations of (a) and (b)  correspond to criterion 1 and WC89 (criterion 5), respectively. The blue points and red crosses represent young and old samples of IRAS sources, respectively. The age is estimated according to a 3D simulation for the expansion of \HII\ regions~\citep{2014A&A...568A...4T}. The positions of blackbodies possessing different temperatures are marked with solid green squares.}
\label{fig:evolve}
\end{figure*}

\subsection{Evolution of \HII\ regions on two-colour diagrams}
\label{sec:evolution}
 
In this section, we examine  the evolution of \HII\ regions on two-colour diagrams. A three-dimensional simulation of the expansion of \HII\ regions performed by  \citet{2014A&A...568A...4T}, involving the effect of the internal turbulence in surrounding gases, enables  us to determine the age of IRAS point sources that are tracing \HII\ regions.

The age calculation for \HII\ regions requires the value of the distance and radio continuum flux. \citet{2014ApJS..212....1A} provides  distances for 1413 \HII\ region candidates, 627 of which are associated with both IRAS and radio continuum sources.  In order to assure the purity of \HII\ regions, we require that the fluxes of all four bands are better than upper limits, and we calculated the ages for the remaining 305 \HII\ regions, based on the results of the 3D simulation of~\citet{2014A&A...568A...4T}.

  The typical age of high-mass stars is about 5 Myr~\citep{2007ARA&A..45..481Z},  and given a timescale of 15\% at embedded phase~\citep{2002ARA&A..40...27C, 2007ARA&A..45..481Z}, the age of \HII\ regions should be substantially less than 5 Myr. However, considering the uncertainties of calculation, for instance, caused by inaccurate assumptions of initial densities of surrounding gases or large errors in distances, we only rejected those \HII\ regions whose ages are greater than 10 Myr, and  adopted those sources younger than 0.5 Myr as the young sample and those sources older than 6 Myr as the evolved sample.   These divisions make the number of young and old samples approximate.

The infrared colours of IRAS point sources tend to be bluer with the evolution of \HII\ regions. We display the result of the colour combinations of criterion 1 (left) and WC89 (right) in Figure~\ref{fig:evolve}.  For both cases, the colours of old samples are generally bluer than that of young samples.  As a comparison, we plotted colours of blackbodies with different temperatures. The difference in the  colour of blackbodies and \HII\ regions indicates that a single blackbody cannot model the whole infrared spectral energy distribution  (SED) of an \HII\ region, and this agrees with the result of~\citet{1999MNRAS.309..905W}, suggesting a two-component blackbody model. This is mainly caused by the distribution of dust grain size~\citep{1976A&A....50..191N,1991AJ....102..658M,2015A&A...576A...2O}, accompanied by the silicate absorption and the polycyclic aromatic hydrocarbon (PAH) emssion~\citep{1999MNRAS.309..905W}.


Nevertheless, the evolutionary trend is consistent with a black body, whose temperature is increasing with time.

 \section{Discussion}
 \label{sec:discussion}

\subsection{Planetary Nebulae}

 Dust surrounding or inside of ionized regions potentially has similar colours even if the ionized gas arises from a PN instead of an \HII\ region.  In order to check this possibility, we matched the \HII\ region candidates (3041) to a Galactic planetary nebulae catalogue~\citep{2013MNRAS.431....2F} which contains 1258 sources. The match radius is 1\arcmin, and in total, we find 22 \HII\ region candidates are near PNe. 

Consequently, less than 1\%   of  \HII\ regions are located nearby PNe, indicating \HII\ regions are well separated from PNe.  20 of those 22 IRAS sources are excluded in the final 2805 \HII\ region candidates identified by criterion1. The left two IRAS sources remain in the candidate list, because they are still possibly tracing \HII\ regions despite neighbouring PNe.


 \subsection{The  evolution of  \HII\ regions}




The evolutionary trend of IRAS sources is consistent with the result of \citet{2003ChJAA...3...49X}. \citet{2003ChJAA...3...49X} investigated 482 6.7 GHz methanol masers, 361 of which are associated with IRAS sources, and they find that on the two-colour diagram, most of those IRAS sources concentrate in a small area, $0.57 \lid \mathrm{log}(F_{25}/F_{12}) \lid  1.30$ and $1.30 \lid \mathrm{log}(F_{60}/F_{12}) \lid 2.5$. They suggest that the infrared colours of  UC \HII\ regions move toward blue colours, which is consistent with the trending revealed by our results, as illustrated in the right panel of Figure~\ref{fig:evolve}.

 \subsection{WC89 and HM89}
  

 As shown in Figure~\ref{fig:criterion5} and Figure~\ref{fig:evowc89},  the criterion of WC89  have missed many evolved \HII\ regions. This is because the  known \HII\ region samples used by WC89 are  UC \HII\ regions, making WC89  sensitive to UC \HII\ regions.  About  half ($\sim$900) of UC \HII\ region candidates selected by WC89 have counterparts in the \wise\ \HII\ region catalogue, meaning that the \wise\ \HII\ region catalogue indeed has missed some UC \HII\ regions.

 As mentioned in \S\ref{totalNumber}, 622 IRAS sources selected by WC89 disagree with criterion 1. However, this does not mean criterion 1 is not compatible with WC89, because most (544) of those 622 sources, having $Q_{100}=1$, are not tested by the inequality of criterion 1. We conclude that the data quality of IRAS PSC may have affected the completeness of \HII\ region candidates. Consequently, those 622 IRAS sources are counted in the total number of \HII\ region candidates in the Milky Way.


As illustrated in Figure~\ref{fig:criterion6}, the area bounded by the decision boundary of HM89 includes a large area which is  dominated by non-\HII\ regions. The decision boundary given by HM89 is $F_{60}/F_{25}\gid 0$ and $F_{25}/F_{12}\gid 0$,   resulted in the selection of many sources at high Galactic latitudes. However, extra constraints on  $F_{100}$ and the Galactic latitudes may have compensated for this inaccuracy. 

 In Figure \ref{fig:venn},  criterion 1 independently identities 1033 IRAS sources,  659 of which disagree with HM89. Those 659 IRAS sources are those \HII\ region candidates  missed by HM89, WC89, and the \wise\ \HII\ region catalogue. 


\section{Conclusions}
\label{sec:conclusions}

We present new criteria for identifying \HII\ regions from the IRAS PSC,  and the criteria are determined by the distribution of  two samples. One  sample (\HII\ regions) is produced by matching the \wise\ \HII\ region catalogue to the IRAS PSC; the other sample (non-\HII\ regions) is constructed by filtering IRAS sources at high Galactic latitudes ($|b|>8\degr$). We determined the decision boundary of the two samples using SVMs, which are efficient classifiers. 

 We find that the optimal selection criterion is  $\mathrm{log}\left(F_{60}/F_{12}\right)\gid \left(  -0.19  \times \mathrm{log}\left(F_{100}/F_{25}\right)+   1.52 \right)$, $Q_{60}> 1$, and  $Q_{100}> 1$, identifying  3041 \HII\ region candidates from IRAS sources,  $\sim$660 of which are new. The known \HII\ regions in our method are  more complete  than that used in HM89, and  we have improved the criterion of HM89 significantly. A high proportion of IRAS \HII\ region candidates have radio counterparts, providing a high confidence for those IRAS \HII\ region candidates. Combining with the \wise\ \HII\ region catalogue, we find that the lower limit  of \HII\ regions in the Milky Way is $\sim$10200.
 
 
We estimate the age of some \HII\ regions based on a 3D simulation involving internal turbulence of surrounding gases, and find that   on two-colour diagrams, younger \HII\ regions have redder IRAS colours and with the evolution, their infrared colours become bluer.

The SVM is an efficient classifier to identify \HII\ regions from infrared surveys. With our methodology and future   sensitive infrared (Herschel, for instance) and radio (recombination lines) observations, we will be able to confirm those \HII\ region candidates and detect more \HII\ regions, approaching the completeness of \HII\ regions in the Milky Way.

\section*{Acknowledgements}

We would like to thank an anonymous reviewer and J. R. Dawson for careful proofreading of the manuscript and constructive comments. This work was partly sponsored by the 100 Talents Project of the Chinese Academy of Sciences, the National Science Foundation of China (Grant Numbers: 11673066, 11233007, 11673051, and 11590781), the Natural Science Foundation of Shanghai under grant 15ZR1446900, and the Key Laboratory for Radio Astronomy.













\bsp	
\label{lastpage}
\end{document}